\documentclass[preprint,12pt]{elsarticle}

\usepackage{amsmath,amssymb,amsfonts}
\usepackage{algorithmic}
\usepackage{algorithm2e}
\usepackage{tabto}
\RestyleAlgo{ruled}
\usepackage{graphicx}
\usepackage{textcomp}

\usepackage{blindtext}
\usepackage{amsthm}
\usepackage{float}
\usepackage{graphicx}
\usepackage{caption}
\usepackage{enumitem}
\usepackage{bbold}
\usepackage{soul,xcolor}
\usepackage{geometry}
\usepackage{multirow}
\usepackage{epsfig}
\usepackage{amscd}
\usepackage{color}
\usepackage{subcaption}
\usepackage{float}
\usepackage[colorinlistoftodos]{todonotes}

\usepackage{mathtools, cuted}
\usepackage{lipsum, color}

\newlength{\arrayrulewidthOriginal}
\newcommand{\Cline}[2]{%
  \noalign{\global\setlength{\arrayrulewidthOriginal}{\arrayrulewidth}}%
  \noalign{\global\setlength{\arrayrulewidth}{#1}}\cline{#2}%
  \noalign{\global\setlength{\arrayrulewidth}{\arrayrulewidthOriginal}}}

\definecolor{darkred}{rgb}{.7,0,0}

\definecolor{darkgreen}{rgb}{0,0.7,0}

\definecolor{darkblue}{rgb}{0,0,0.7}

\numberwithin{equation}{section}
\numberwithin{figure}{section}
\numberwithin{table}{section}

\begin{document}

\begin{frontmatter}

\title{Interpretable Forecasting of Physiology in the ICU Using Constrained Data Assimilation and Electronic Health Record Data}

\author[dbmi,bioe,c,d]{David Albers}
\author[dbmi]{Melike Sirlanci}
\author[e]{Matthew Levine}
\author[f]{Jan Claassen}
\author[g]{Caroline Der Nigoghossian}
\author[d]{George Hripcsak}

\address[dbmi]{{Department of Biomedical Informatics},
            {University of Colorado Anschutz Medical Campus},
            {Aurora},
            {80045},
            {CO},
            {USA}}

\address[bioe]{{Department of Biomedical Engineering},
            {University of Colorado Anschutz Medical Campus},
            {Aurora},
            {80045},
            {CO},
            {USA}}

\address[c]{{Department of Biostatistics and Informatics},
            {University of Colorado Anschutz Medical Campus},
            {Aurora},
            {80045},
            {CO},
            {USA}}

\address[d]{{Department of Biomedical Informatics},
            {Columbia University},
            {New York},
            {10032},
            {NY},
            {USA}}

\address[e]{{Department of Computing and Mathematical Sciences},
            {California Institute of Technology},
            {Pasadena},
            {91125},
            {CA},
            {USA}}

\address[f]{{Division of Critical Care Neurology, Department of Neurology},
            {Columbia University},
            {New York},
            {10032},
            {NY},
            {USA}}

\address[g]{{School of Nursing},
            {Columbia University},
            {New York},
            {10032},
            {NY},
            {USA}}

\begin{abstract}
Prediction of physiologic states are important in medical practice because interventions are guided by predicted impacts of interventions. But prediction is difficult in medicine because the generating system is complex and difficult to understand from data alone, and the data are sparse relative to the complexity of the generating processes due to human costs of data collection. Computational machinery can potentially make prediction more accurate, but, working within the constraints of realistic clinical data makes robust inference difficult because the data are sparse, noisy and nonstationary. This paper focuses on prediction given sparse, non-stationary, electronic health record data in the intensive care unit (ICU) using data assimilation, a broad collection of methods that pairs mechanistic models with inference machinery such as the Kalman filter. We find that to make inference with sparse clinical data accurate and robust requires advancements beyond standard DA methods combined with additional machine learning methods. Specifically, we show that combining the newly developed constrained ensemble Kalman filter with machine learning methods can produce substantial gains in robustness and accuracy while minimizing the data requirements. We also identify limitations of Kalman filtering methods that lead to new problems to be overcome to make inference feasible in clinical settings using realistic clinical data.
\end{abstract}

\begin{graphicalabstract}
\includegraphics[scale=0.5]{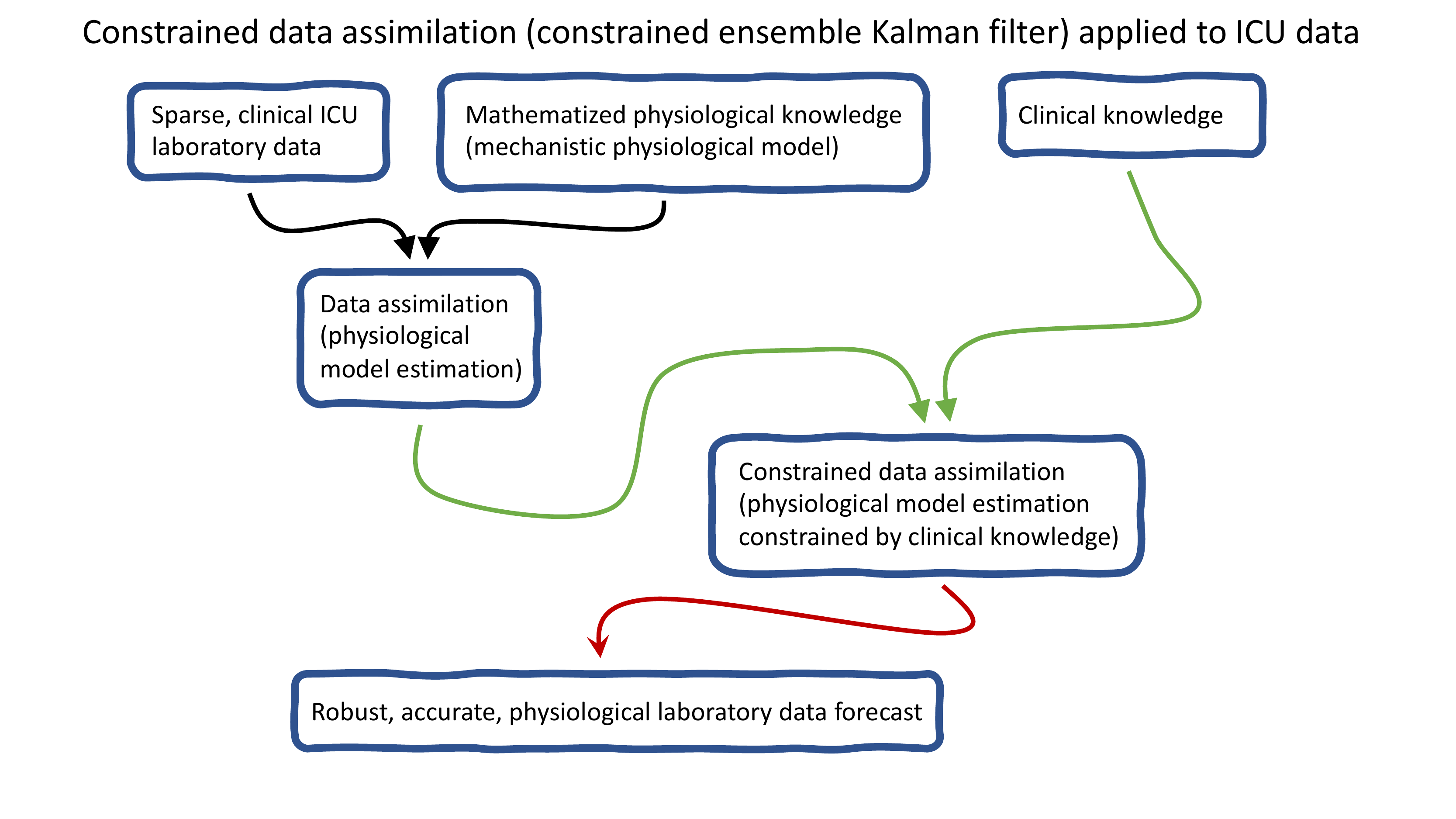}
\end{graphicalabstract}

\begin{highlights}
\item We apply data assimilation (DA) to forecast the glucose-insulin system of ICU patients.

\item We introduce a new method of constrained DA that incorporates clinical and physiological knowledge into a  DA framework to forecast endocrine physiological mechanics using ICU data.

\item The new DA method reduces mean-squared error of forecasting error while requiring a fraction of the data, but the reduction in MSE can come with induced and unintended physiological errors as a fundamental property of MSE minimization.


\item While we perform the analysis on the glucose-insulin system, the results generalize to any sparsely measured physiological system estimated with clinical data.

\end{highlights}


\end{frontmatter}

\section{Introduction}


Practicing medicine requires continual and accurate prediction: clinicians undertake interventions they predict will help the patient. But prediction is complex in medicine. Identifying the presence of cancer from a biopsy requires making an observation. Then, when this observation is coupled with further aggregate information from laboratory results, computational omics-related results, and other patients via, e.g., clinical trials, the synthesis of this patient-specific observation with prior information leads to a prediction about how to best care for the patient. Through this process of synthesizing information, medical care is personalized. Meaning, embedded in the medical personalized care approach is an assumed goal, observations and the information content within patient data, and prior knowledge derived from years of research. In this way, to make medicine personalized, the clinician must rely on limited, often noisy, data from an individual, together with pre-existing imperfect prior knowledge. Synthesizing personal data and prior knowledge is combined to make a prediction regarding care. The more information that can be extracted from the existing data, the more personalized the predictions and treatments can be.


\paragraph{\textbf{Data sparsity and physiological prediction with clinical data}} One of the most vexing problems for personalizing medical care---both human and machine-based---is data sparsity. \textbf{The primary focus of this paper is estimating and forecasting endocrine physiological mechanics, specifically of the glucose-insulin system, given sparse clinical data.} This system is used as a testbed to explore and identify problems because it is a representative of many other situations. Personal clinical data are often sparse relative to the complexity of the system generating the data, but sparsity is not a precisely defined concept. \emph{Here we conceptualize data sparsity by the amount of information in the data relative to the information required for a prediction task.} Sometimes, the same data can be sparse for one prediction task and dense for another. For example, given the task of identifying the presence of cancer, the information in a biopsy is generally not sparse and can be used to identify a cancer assuming previous knowledge of the cancer. In contrast, given the task of understanding the cancer, its source, its nature, how it works, and how to stop it, the information in a single biopsy is incomplete. Sparse data makes prediction difficult because sparsity of information in the data alone makes it difficult to \emph{uniquely and accurately} select a predictive model that will be representative of the system generating the data.


\paragraph{\textbf{Data assimilation as an approach to managing data sparsity} } Our approach to coping with data sparsity begins with data assimilation (DA), a broad class of methods including statistical and variational methods for estimating states, parameters, and initial conditions of mechanistic mathematical models \cite{stuart_da,filtering_jaz,da_siam}. DA methods are usually conceptualized as a pairing of a mechanistic model with an inference method because inference methods are usually applied in the context of a particular mechanistic model. Here the model is a mathematical physiological model. Estimation accuracy depends on both the model, the inference method \cite{albers_plos_comp_bio_DA_I}, and data. DA copes with data sparsity by augmenting the sparse data with knowledge of biophysical mechanisms through the use of a mathematical model. We developed a DA method, based on the ensemble Kalman filter (EnKF), that incorporates additional knowledge of the underlying generating system in the form of constraints on states and parameters that, e.g., allow us to set limits on insulin levels the model can have. Our constrained DA method was motivated by data sparsity problems encountered when using clinical data to make personalized predictions.



\paragraph{\textbf{The primary foci of the paper}} This paper has two foci within the context of using DA for forecasting physiological processes with sparse and non-uniformly sampled (intensive care unit) ICU time series of \emph{$10-20$} measurements. DA, being a broad class of methods, has many realizations. We focus on the filtering and forecasting method classes \cite{filtering_jaz,da_siam,stuart_da} that most commonly are variations of Kalman filters. Recalling the standard computational workflow for DA in these contexts where the model and inference method choices have been made, we: select which model parameters to estimate (we always estimate the model states), set initial conditions of the ODE model parameters and states, begin simulating the model, estimate the model states and parameters when measurements arrive, synchronizing the model with the observed system states and estimating the underlying physiological parameters. \textbf{Within this context the paper has two primary foci.} \emph{First}, we demonstrate a new DA method, the constrained Ensemble Kalman Filter (CEnKF) that we previously proved \cite{cenkf} was optimal in the sense that it minimizes least squares error in the situation where forecasts just remain within constrained regions, using the data the method was motivated to solve. This method reduces forecasting errors substantially, but reveals new and unexpected sparse-data problems that remain even when we apply our DA method that includes constraints based on clinical and physiological understanding. This raises the \emph{second focus}, we observe and quantify a fundamental problem estimating a model with sparse data that is embedded in the cost or error function being minimized. Moreover we find this problem is particularly severe when the sparse data correspond to an oscillatory generating function such as the case with the glucose-insulin system under constant nutrition infusion \cite{sturis_91}. Specifically, minimizing least squares ($L_2$) errors between the model and data yield a very good---\emph{potentially optimal but wrong}---solution as the constant conditional mean or a conditional mean whose graph lies between the minimal and maximum of observations. For a machine learning system, this solution may be acceptable if uninformative clinically, but for the mechanistic models embedded in DA where parameters have physiological meanings, parameters that lead to model dynamics that are constant fixed points rather than oscillatory periodic orbits represent the wrong underlying physiological mechanics.



\paragraph{\textbf{Problem background}} To understand the motivation of these two problems and how they arose, consider three fundamental problems when applying DA to clinical data to predict physiological systems. \emph{First}, models must be personalized to be accurate. We know from previous work that without parameter estimation, models fail to accurately represent patients \cite{albers_plos_comp_bio_DA_I,levine_albers_plos_comp_bio_DA_II,jamia_da,clermont_inverse}. \emph{Second}, identifiability failure (or failure to have unique solutions) is common because of data sparsity, a problem that can be managed but never eliminated given realistic clinical data. To reduce this problem we have four options: \emph{(a)} we can define simple identifiable models \cite{melike2019simple}, \emph{(b)} restrict parameter estimation such that models become more identifiable \cite{houlihan}, \emph{(c)} constrain inference \cite{cenkf} to mitigate identifiability failure and \emph{(d)} apply both (b) and (c), a situation that is often necessary because unmeasured states may be important modeling components and are not amenable to implicit incorporation into the model. \emph{Third}, we must ensure we \emph{infer parameters that reproduce the correct model dynamics}, e.g., the topology or invariant measure, of the observed system. If the underlying system is oscillating but the model has no oscillations, the model parameters correspond to qualitatively different physiological mechanics than that of the observed system. \emph{Returning to the two results in this paper,} we show how and why it is often necessary to constrain inference to make accurate predictions when using real clinical data. Moreover, this paper shows how to solve, and evaluates the solution to, the constrained inference problem with ensemble Kalman filters in the context of real ICU clinical data---i.e., the first focus point of the paper. But, constraining inference when data are sparse can lead to a surprising consequence: the error minimizing solutions to the constrained inference problem do not necessarily lead to parameter sets that produce the correct generating dynamics. In fact, the constraints may actively lead to solutions with lower errors than the unconstrained inference whose parameters can produce the wrong topology and invariant measure when data are sparse---i.e., the second focus of the paper. In opposition, when constraints are not imposed, the inference rarely converges due to the sparsity of data. These findings together demonstrate a substantial gap in our ability to use real clinical data for personalized prediction and are the primary take away of this paper.


\paragraph{\textbf{Clinical context} }We demonstrate these results in the clinical context of the ICU, using glycemic management as the prototype case. We apply DA methods to forecast and estimate BG-related physiology in an ICU, and therefore restrict the use of data to those collected according to standard ICU practice---real world evidence \cite{real_world_evidence}---treating the data as if they arrive prospectively. ICU glycemic management approximately adheres to the following algorithm. Clinicians maximize caloric intake because patients rarely consume enough nutrition. Given this, clinicians measure and manage glucose levels through insulin administration specified by a non-personalized flowchart \cite{51_protocols}. As many as $90 \%$ of patients receive insulin despite not needing it outside the ICU setting and $20 \%$ of patients have a hypoglycemic event requiring emergency glucose administration. Glycemic need depends on the disease process and therefore varies by ICU \cite{36_intenstive_insulin_therapy,37_NICE,42_mike_schmidt_metabolism_SAH,43_tight_NICU_glucose,44_glucose_microdialysis_NICU_other,45_NICU_functional_outcome_glycemic_management}. Glycemic management has an impact on morbidity and mortality \cite{32_glucose_control_ICU,33_hyperglycemia_review} and is time consuming for nurses to manage \cite{46_NICU_hard_to_reach_glycemic_targets}. And even with extensive effort, such as during clinical trials, glycemic targets are difficult to meet \cite{30_how_tight_too_tight}. Nevertheless, there does seem to be significant potential to improve outcomes through glycemic management \cite{37_NICE}. There have been efforts to use closed-loop control in the ICU to mange glucose \cite{22_leearathna,23_lin2004adaptive,24_lin2011physiological,25_knab2015zone,26_knab2016virtual,27_haverbeke2008nonlinear,28_vanherpe07,closed_loop_glucose_control}, none seem to have succeeded for complex patients or when interventions or patient state changes, and the likely reason seems to be correctly estimating the glycemic response to insulin administration \cite{29_clermont_pritchard2017modeling} as well as the impact of the myriad of drug-based therapies, e.g., glucocorticoids, that impact glycemic homeostasis. When patients are stationary---in homeostasis---they are relatively easy to predict, but in this case, sophisticated prediction is not needed. Instead it is the deviations from homeostasis that cause the operational problems. We hypothesize that this problem is rooted in a mix of model error, e.g., medications such as glucocorticoids are not represented explicitly as states or parameters in the models, and identifiability failure, the later of which we address in this paper. It is hopeful that continuous glucose monitors (CGM) will become more common in ICUs, especially post-COVID-19 \cite{icu_covid_cgm,covid_icu_poc_protocol} and CGM may mitigate some of these problems, but clearly not all of them, e.g., insulin will remain unmeasured. And, in any event, the data sparsity problem will not, in general and over the broader ICU context, vanish in the near future.

Within this context, the computational problem for DA is to generate accurate state (e.g., glucose) forecasts, parameter (e.g., insulin secretion) estimates, and forecast uncertainty, in real-time. In this case, glycemic management in the ICU requires inference that converges to an accurate model within $1$ day or $<24$ data points, ideally within 8 hours or 8 data points---and the forecast accuracy must persist as the patient evolves due to interventions and improvement. Adding CGM data would increase the number of glucose measurements and likely aid the DA machinery in this context, but this sparse data problem is so prevalent in clinical environments that the primary points of the paper will generalize widely beyond glucose forecasting.

\noindent
\textbf{Statement of significance:}

\noindent
\textbf{Problem:} Forecasting physiological features using clinically collected data is challenging because of data sparsity where models are estimated, and predictions are made with $10-20$ data points.

\noindent
\textbf{What is already known:} One way to address the data sparsity problem is with data assimilation (DA) where a mechanistic mathematical model of physiological mechanics is estimated and used to forecast. However, DA can encounter errors because many important features, e.g., insulin in the glucose-insulin system, are not measured.

\noindent
\textbf{What this paper adds:} This paper introduces a new constrained DA method, the constrained ensemble Kalman filter, that injects clinical and physiological understanding in the form of model constraints. The method reduces mean-squared error of forecasting error while requiring a fraction of the data, but the reduction in MSE can come with induced and unintended physiological errors as a fundamental property of MSE minimization.

\section{Methods}

\subsection{Data}
We are applying DA methods to forecast and estimate BG-related physiology in an ICU, and therefore restrict the use of data to those collected according to standard ICU practice---real world evidence \cite{real_world_evidence}---treating the data as if they arrive prospectively. The data available include: blood glucose (BG) measurements---finger stick and IV measurements roughly once an hour or less, insulin administration including rapid and short acting IV drip and injectable insulin (we exclude intermediate-acting, long acting.), and nutrition including tube feed, incidental IV drip that may contain glucose, and emergency IV glycemic bolus used to correct for hypoglycemia. \emph{Plasma insulin is effectively never measured.}

The de-identified medical records were collected retrospectively from nine patients from the Neurological Intensive Care Unit at the Columbia University Irving Medical Center using the Clinical Data Warehouse; their collection and use was approved by the institutional review board. The data were curated by hand to ensure their accuracy. In table ~\ref{table:data_summary}, we report the types of measurements and treatments known for each patient, along with frequencies and severity of glucose measurements. Our data do not include CGM data as they are currently rare in an ICU setting, a situation that may change in the future \cite{icu_covid_cgm,covid_icu_poc_protocol}. Similarly, there are no insulin measurements because it was never measured in the ICU, as is generally the case.


\begin{table*}
\centering
\resizebox{\textwidth}{!}{
{
\begin{tabular}{|l| |l| |l| |l| |l| |l| |l| |l| |l|}
  \hline
  \multicolumn{9}{|c|}{ Data summary table} \\
  \hline \hline
  id & $\#$ of GLU  & days & GLU $\# < 70, >40 $ &  $ GLU \#  < 40$ & insulin & IV GLU drip & IV GLU bolus & Tube feed   \\ \hline \hline
296  & 230 & 19 & 3 & 0 & 6 & 228 &  0  & 253 \\ \hline
426  & 177 & 14 & 0 & 0 & 40 & 141 &  0 & 289 \\ \hline
430  & 227 & 22 & 1 &  0 & 26 & 400 &  0 & 216 \\ \hline
456 &131  & 22 & 0 & 0 & 48 & 46 &  0 & 176 \\ \hline
489 & 104 & 13  & 1 &  0 & 22 & 260 &  2 & 236 \\ \hline
585 & 405 & 78  & 3 & 0 & 30 & 389 &  1 & 1441 \\ \hline
593  & 262  & 27 & 2 & 0 & 4 & 60 & 0  & 251 \\ \hline
646 & 153 & 19 & 0 & 0 & 51 & 49 &  0 & 14 \\ \hline
851 & 289 & 26 & 4 & 1 & 125 & 51 &  2 & 301 \\ \hline
  \hline \hline
\end{tabular}
}
}
 \caption{Summary of the data collected per patient and made available to the DA, including the ID, number of glucose measurements, number of glucose measurements less than 70 but greater than 40, number of glucose measurements less than 40, number of insulin deliveries, number of glucose drip nutrition delivery instances, number of glucose IV boluses given, number of tube feed instances. Glucose measurements below 40 usually triggers a bolus delivery, and glucose below 70 triggers an urgent intervention. Tube feed, glucose IV drip, and glycemic bolus instances often indicate changes to otherwise continuous infusions of nutrition or insulin.}
   \label{table:data_summary}
\end{table*}

\subsection{Data assimilation}
\label{sec:da}

Data assimilation is a mathematical field that provides a framework for combining mathematical models and observational data to achieve the desired goals in an optimal manner  \cite{stuart_da,filtering_jaz,da_siam}. In order to perform state estimation, forecasting, or control, we need mathematical models that describe the system of interest. However, as these models are only an approximation to  reality, unless corrected or modified regularly based on data, their representation of reality substantially degrades. This is exactly where data assimilation comes into play, due to its ability to synthesize a theoretical (mathematical) model and data in a possibly optimal way.

Here, we have considered the model presented by Sturis \emph{et al.} \cite{sturis_91}, whose equations can be summarized as $\dot{v} = f(v,t)$, where $v$ is the state space of the model (i.e. glucose and insulin concentrations), and $f$ represents the time-dependent dynamics of the model (i.e. the right-hand-side of the system of differential equations). We can then write a solution for this system at time $t$ with initial state $v_0$ as $v(t,v_0)$.


More generally, we can let $\Psi_j$ be the solution operator that maps the continuous-time system from a state $v_j$ to the next state $v_{j+1}$. We write:
$$v_{j+1} = \Psi_j(v_j) := v(t_{j+1}, v_j).$$


Note that the dependence of $\Psi$ on $j$ allows for non-autonomous $f(v,t)$ and non-uniformly spaced times.

Although underlying state dynamics are governed by the continuous $f(v,t)$, we can use $\Psi_j$ to integrate $f$ between measurement time points (often with a numerical solver) and represent the evolution of observations as a discrete process.
This formulation fits nicely into the general data assimilation framework provided by Stuart \emph{et al.} \cite{stuart_da}, which considers a discrete-time dynamical system with noisy state transitions and noisy observations:
\begin{align*}
	\text{Dynamics Model:}  \quad v_{j+1} &= \Psi_j(v_j) + \xi_j, \quad j \in \mathbb{Z}^+ \\
	\text{Data Model:}  \quad y_{j+1} &= Hv_{j+1} + \eta_{j+1}, \quad j \in \mathbb{Z}^+ \\
	\text{Probabilistic Structure:}  \quad v_0 &\sim N(m_0, C_0), \quad \xi_j \sim N(0, \Sigma), \quad \eta_j \sim N(0, \Gamma)\\
 \text{Probabilistic Structure:} \quad v_0 &\perp \{\xi_j\} \perp \{\eta_j\} \text{ independent}
\end{align*}
where $\xi_j \sim \mathcal{N}(0,\Sigma), \eta_{j} ~\sim \mathcal{N}(0, \Gamma)$ are assumed noises in the model state dynamics and measurement operations with mean zero and covariances $\Sigma$ and $\Gamma$, respectively. We also assume a known distribution $N(m_0, C_0)$ for initial condition $v_0$.
$H$ is a linear measurement operator, which in our case selects only the glucose component of the full physiologic state vector $v_j$ to produce a glucose measurement $y_j$. We collect $M$ glucose measurements $\{y_j\}_{j=1}^M$ at times $\{t_j\}_{j=1}^M$.

We are primarily interested in leveraging historical data $\{y_j\}_{j=1}^J$ to improve our understanding of the current system state $v_J$ such that our forecast of $v_{J+1}$ is optimized. We do this by solving a filtering (state estimation) problem, where a sequential estimation scheme is used to first estimate the distribution of $v_j$ given $y_j$ (i.e. assimilate the data), then forecast $v_{j+1}$ using this assimilation.

For DA purposes, there are a wide variety of powerful algorithms that could be used depending on the structure of the mathematical model. When the dynamics model, $\Psi(\cdot)$ is linear, then it is possible to use a Kalman Filter. However, when $\Psi(\cdot)$ is not a linear map, we can no longer use Kalman Filtering and need other algorithms to handle the nonlinearity of $\Psi(\cdot)$. In this case, one method is the Extended Kalman Filter, which applies the Kalman Filter methodology, using the linearized version of the map $\Psi(\cdot)$ to update covariances. Even though this model is useful in addressing the nonlinearity, it is not computationally convenient when the system is high-dimensional, since it requires linearization and exact covariance computation at every step. In this case, one could use an Ensemble Kalman Filter (EnKF) as it provides a computationally efficient covariance update at every step. Therefore, since it addresses all the requirements of our model, we decided to use the EnKF algorithm by introducing constraints into the original algorithm. Both the unconstrained (original) and constrained versions of the EnKF are described in detail in the following two sections, respectively.

\textbf{Ensemble Kalman filter:}
The ensemble Kalman filter performs state estimation by sequentially updating an ensemble of particles to achieve an optimal balance between observed data and model-based predictions.
We create particles $\{{v}_{j}^{(n)}\}^{N}_{n=1}$ to represent $N$ potential state estimates at time $j$.
We use the dynamics model $\Psi$ to make predictions for each particle at time $j+1$
to give $\{\widehat{v}_{j+1}^{(n)}\}^{N}_{n=1}$, whose statistics reflect our probabilistic prediction about the state at time $j+1$.
We use the resulting empirical ensemble covariance
to define the objective function $I_{{\rm filter},j,n}(v)$, which specifies a compromise between trust of the model and the data; this tradeoff characterizes our data-driven update step.
We minimize the objective function to obtain the updates $\{{v}_{j+1}^{(n)}\}^{N}_{n=1}.$

\textbf{Prediction:} The following prediction step produces an estimate of the mean ($\widehat{m}_{j+1}$) and covariance ($\widehat{C}_{j+1}$) of the physiologic system at time $j+1$
\begin{subequations}
\label{eq:deenz}
\begin{align}
\widehat{v}_{j+1}^{(n)} &= \Psi(v_{j}^{(n)})+\xi^{(n)}_{j}, n=1,...,N \\
\widehat{m}_{j+1} &= \frac{1}{N}\sum^{N}_{n=1} \widehat{v}_{j+1}^{(n)} \\
\widehat{C}_{j+1} &= \frac{1}{N}\sum^{N}_{n=1}\bigl(\widehat{v}^{(n)}_{j+1}-\widehat{m}_{j+1}\bigr)\bigl(\widehat{v}^{(n)}_{j+1}-\widehat{m}_{j+1}\bigr)^{T}. \label{eq2pt1c}
\end{align}
\label{eq:this}
\end{subequations}
with $\xi_{j}^{(n)} \sim N(0,\Sigma)$ i.i.d..

\textbf{Update:} The update step is
\begin{equation}
v_{j+1}^{(n)}=\underset{v}{\mathrm{argmin}}\,I_{\rm{filter},j,n}(v), n=1,...,N,
\end{equation}
where
\begin{equation}
\label{eq:deenz2}
I_{{\rm filter},j,n}(v) :=
		\frac{1}{2} \mid y_{j+1}^{(n)} - Hv \mid ^{2}_{\Gamma} + \frac{1}{2} \mid v-\widehat{v}_{j+1}^{(n)} \mid ^{2}_{\widehat{C}_{j+1}}, \text{ for each } n=1,...,N,
\end{equation}
with $|v|_B^2  = v^* B^{-1} v $.

Solutions of the minimization problem \eqref{eq:deenz2} can be obtained through
the well-known Kalman update formulae (see \cite{stuart_da} for a derivation):
\begin{subequations}
\label{eq:update}
\begin{align}
S_{j+1} &= H\widehat{C}_{j+1}H^{T} + \Gamma \\
K_{j+1} &=\widehat{C}_{j+1}H^{T}S_{j+1}^{-1} \qquad(\rm{Kalman}\,{\rm Gain}) \\
v_{j+1}^{(n)} &= (I-K_{j+1}H)\widehat{v}_{j+1}^{(n)}+K_{j+1}y_{j+1}, n=1,...,N
\end{align}
\end{subequations}

\textbf{Constrained Update:} As demonstrated in \cite{cenkf}, the filter can be constrained to a pre-defined state and parameter space by performing the optimization in \eqref{eq:deenz2} with respect to linear equality and inequality constraints on $v$. Thus, the EnKF remains well-defined for any choice of $A,B,a,b$ defining a constraint set:

\begin{align*}
Av &= a\\
Bv &\leq b
\end{align*}

In practice, this means that we can impose upper and lower bounds for states and parameters of our physiologic model, and ensure that all updated particles in the filter obey these constraints.

\subsection{Physiologic endocrine models}

\subsubsection{Model choice}

We use a model consisting of a set of ordinary differential equations (ODEs) to represent glucose-insulin dynamics, the Ultradian model \cite{sturis_91}, which models the interaction of glucose, plasma insulin and interstitial insulin, together with a three-stage delay filter modeling hepatic (liver) glucose production, and which parameterizes all other processes such as insulin secretion. The linear filter is an implementation of the linear chain trick that allows the use of an ordinary differential equation to approximate a time-delayed differential equation. We chose this model because it was not created for a particular setting, it resolves ultradian oscillations and is well known and validated. Ultradian oscillations are important in the ICU setting because many critical care patients are tube-fed and continuous feeding leads to oscillatory glycemic dynamics rather than the more typical damped-driven oscillatory dynamics resulting from meals \cite{pop_phys}.  In other settings such as type-2 diabetes we have found the ultradian model performs well within the type-2 diabetes setting as well \cite{albers_plos_comp_bio_DA_I,melike2019simple}, meaning the ultradian model has generalizable utility. There are models that have been constructed explicitly for particular settings such as the ICU minimal model \cite{vanherpe06} or the glucose and subcutaneous insulin model \cite{29_clermont_pritchard2017modeling} in the ICU or models that focus on diabetes \cite{corbelli_review} and its progression \cite{sherman_beta_cell_model_I,artie_ha_ldpm_II}. Because we focused on inference here, and because the ultradian model is able to represent oscillatory glycemic dynamics in the ICU, we choose to estimate the ultradian model and  avoid comparing the ICU-specific glucose models. The problems we investigate here will be present in any model that includes insulin and oscillatory BG dynamics. The model details are presented in the appendices.

\subsubsection{Ultradian model parameter choice} The Ultradian model has a higher degree of biophysical accuracy than we can resolve with data and as a result we cannot estimate most of its parameters. This problem arises because ultradian model parameters are coupled, redundant and estimating all parameters induces severe identifiability issues.  Nevertheless, it is important to understand and control for model error induced by estimating a limited parameter set--when the model is not tailored to an individual, the model accuracy suffers substantially \cite{levine_albers_plos_comp_bio_DA_II}.  As such we used the Houlihan parameter selection method \cite{houlihan} tuned to be sensitive to mean and the variance of glucose to select two subsets of parameters, one set with five and one set with eight parameters respectively. Houlihan methodology as it applies to the ultradian model is detailed in \cite{houlihan}, and the parameter sets we selected here were roughly taken from that work. The maximal eight-parameter set includes $(R_g, C_3, U_m, a_1, C_1, t_p, R_m, t_d)$; this is the entire set of parameters selected by the LASSO-based Houlihan with the exception that we excluded $V_p$, the parameter that adjusted the volume of plasma and added $t_d$, the parameter that controls the hepatic delay between plasma insulin and glucose production. We excluded the volume parameter because the volume of humans in our data set is not particularly variable; this parameter was the least important parameter according to the Houlihan so its exclusion is not particularly impactful.  We included the parameter that controls the delay between plasma insulin and glucose production because dynamically it controls the delay term that controls how and whether glucose and insulin oscillate under constant nutrition input, a feature we know is present from prior work of Sturis et al \cite{sturis_91,sturis_chaos}. The minimal parameter set is a subset of the maximal set, including $(R_g, C_3, a_1, C_1, t_d)$ chosen in a way to include the most physiological compartments of the model that can be tuned while reducing the parameter space being estimated by $\frac{3}{8}$. This second parameter set is used to contrast the maximal parameter set in terms of induced model error by limiting the flexibility of the model while still estimating important and diverse parameters.

\subsection{Forecasting experiments}

\subsubsection{Overarching computational experimental procedure}

The overarching structure of our computational experimental is shown in Algorithm (\ref{alg:experiment}). For each experiment, we assume a mechanistic physiological model. Over the course of the experiments we vary the the DA method between the EnKF or CEnKF, the choose parameter subset we allow DA to estimate, the number of particles in the ensemble, the constraints on the CEnKF, the data the model can use to forecast blood glucose, and the patient.

\SetKwComment{Comment}{/* }{ */}


\begin{algorithm}
\SetKw{Analysis}{Analysis}
\caption{Computational experimental procedure}\label{alg:experiment}
\KwData{patient glucose, insulin, administered carbohydrates, $n$ blood glucose measurements}
\KwOut{ensemble of continuous-time blood glucose forecasts and corrections, ensemble of parameter estimates at measurement and intervention times, MSE between forecast ensemble mean and measurements}


\If{$i=1$}{
initialize the experiment: select parameter subset for estimation, set number of ensemble particles, set EnKF constraints, set the input data (e.g., include or not incidental IV-dextrose)\;

  }

\While{$i \leq n$}{
	iterated the DA forward to the next measurement\;

\If{$t_i \geq 24$ hr}{
 $MSE(i) = MSE(i-1) + ||y(t_i)-H \hat{m}(t_i)||^2$\;
 \Comment*[r]{MSE between forecast ensemble mean and blood glucose measurement}\;
 }
 return next-measurement prediction MSE\;
 }
\KwResult{scientific analysis of DA output}

Comparison of cumulative MSE with clinical decision requirements\;
Comparison between different error metrics\;
Assessment of time to model convergence\;
Assessment of constraint violation\;
Qualitative comparison of model ensemble errors with data\;
Qualitative comparison of model dynamics and known generating dynamics\;
\end{algorithm}

%
%
%
%
%
%

\subsubsection{Forecasting with the EnKF with no constraints}

\paragraph{Model variation} we investigate how model complexity and parameter estimation choices affect the inference and forecasting of glucose and endocrine state and what consequences these differences have given clinical data.  To vary the model we infer two different parameter sets of the ultradian model, one being a subset of the other.  The parameter sets were selected using a combination of the Houlihan method \cite{houlihan} and knowledge of the model, resulting in the previously discussed data sets, the Houlihan, $P_H$, and the restricted Houlihan, $P_{RH}$ parameter sets.


\paragraph{Parameter estimation} We use the joint EnKF to estimate both states and parameters. Because we use the EnKF in a filtering set-up, it predicts and corrects states and parameters adaptively as data arrive, and because of this the initial parameter values for the filter matters a great deal.  If the initial parameters are near good parameter estimates for a given patient, very little adaptation has to be done to achieve a good forecast.  To control for this, we \emph{always} begin the EnKF at the nominal parameters for the models.  These parameters may be far or close to a patient's ideal parameter estimates but they remain consistent for all the experiments for a given model.

\paragraph{Particle variation} We varied the number of particles in the EnKF between $15$, $50$, and $500$. We present results with N=50 because it was nominally the N that minimized the effects of ensemble size on the inference. Setting N very high or very low can have advantages and disadvantages that help understand why we set N=50. Larger ensembles can capture uncertainty better but also Gaussianify the ensemble more. Moreover, larger ensembles slow the filter down because of the increased number of evaluations of $\Psi(v^{(n)}_j)$. In addition, the tails of the distribution are better represented with more particles, making constraint violation more likely, and thus requiring an optimization of \eqref{eq:deenz2}, rather than the faster matrix inversion steps. We found that using $50$ particles produced the nicest and most fair balance between estimation quality and estimation time. There was not a substantial decrease in error when ensemble sizes were increased to $500$. Nevertheless, ensemble sizes of $15$ particles did surprisingly well.

\subsubsection{Forecasting with the EnKF: nutrition delivery data inclusion experiments}

\paragraph{Data variation} We work with several data constructions, a \emph{simple data} where we include tube nutrition, blood glucose measurements, and all forms of administered insulin and two \emph{more expansive} data sets that include tube nutrition, blood glucose measurements, all forms of administered insulin, and \emph{either} IV glucose bolus \emph{or} incidental IV glucose drips \emph{delivered though the process of administering drugs though $5 \%$ dextrose in water (D5W)}. We include these data permutations because real-time data availability and accuracy are variable: inadvertent IV glucose nutrition, or glucose delivered via dextrose $5 \%$ in water (D5W) over the course of other medication delivery is embedded in medication narrative (written notes) and can be difficult to extract in real-time, tends to be error prone and represents a significant clinical workflow barrier to using DA in an ICU setting. As such, we wanted to understand and evaluate the need of calories delivered via D5W.

\subsubsection{Forecasting with the EnKF with constraints}

Table \ref{table:constraints} shows the boundaries we impose on the UM-CEnKF parameter estimation. We have three categories of constraints: mild, severe, and ultra severe. Note that only insulin has "ultra severe" constraints. This is because insulin levels have a profound influence on glycemic dynamics, yet are not identifiable states because insulin is never directly measured in the ICU setting. The mild constraint bounds on parameters are the nominal parameters adjusted up or down by an order of magnitude for the upper or lower bounds respectively.  The severe constraint bounds on parameters are plus or minus one half of the nominal parameter value for the upper or lower bounds respectively. Glucose and insulin constraints are based on levels of normal variation of glucose and insulin. Based on this, we conduct $11$ constraint experiments detailed in Table \ref{table:constraints} .

\subsection{Evaluation methods}

We primarily use the mean squared error (MSE) between \emph{point-wise forecasts} and data. Because we are evaluating forecasting, the UM-EnKF/UM-CEnKF makes predictions into the future up until the next measurement and so in some sense every data point represents a new and independent evaluation of the UM-EnKF/UM-CEnKF  forecast. The UM-EnKF/UM-CEnKF  is continuously estimating parameters, so the forecast evaluation method, while standard in an online UM-EnKF/UM-CEnKF  context, carries a different interpretation compared to smoothing or inverse problems contexts. We begin to calculate the MSE $24$ hours after admission to the ICU, which implies a variable set of data points usually $<24$ and nominally in the range of $12-18$ data points. We report the MSE after $24$ hours because we would need the DA-based decision support system to be accurate no later than $24$ hours from admission. In this setting, we need the UM-EnKF/UM-CEnKF  to issue a forecast within one day (or sooner) of the patient arriving in the ICU, and we do not have the benefit of more data or information than is available in the first $24$ hours. The MSE is not the best MSE we can achieve, but rather reflects a conservative average operational MSE we might see in a real clinical setting. Moreover, to provide a deeper evaluation we will show entire MSE curves from start to end and this will help give context and intuition to the consequences of this choice. We will also use the glucose distributions, probability densities of forecasts and measurements, to qualitatively evaluate the forecasts.  We will not use these methods quantitatively here.

\section{Results}


\subsection{Unconstrained, EnKF-based forecasting and estimation}

\begin{figure*}
\includegraphics[scale=0.17]{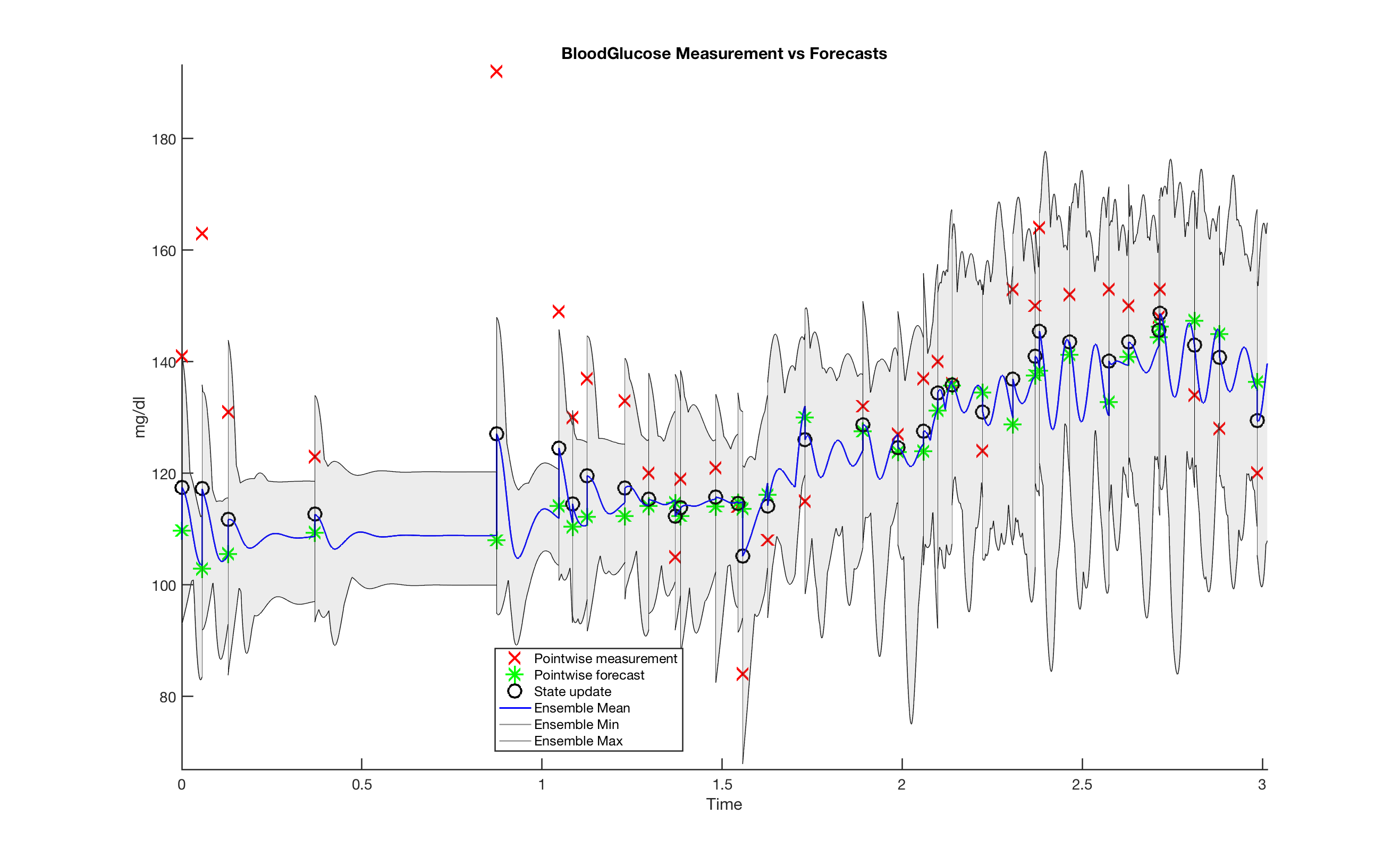}
\includegraphics[scale=0.17]{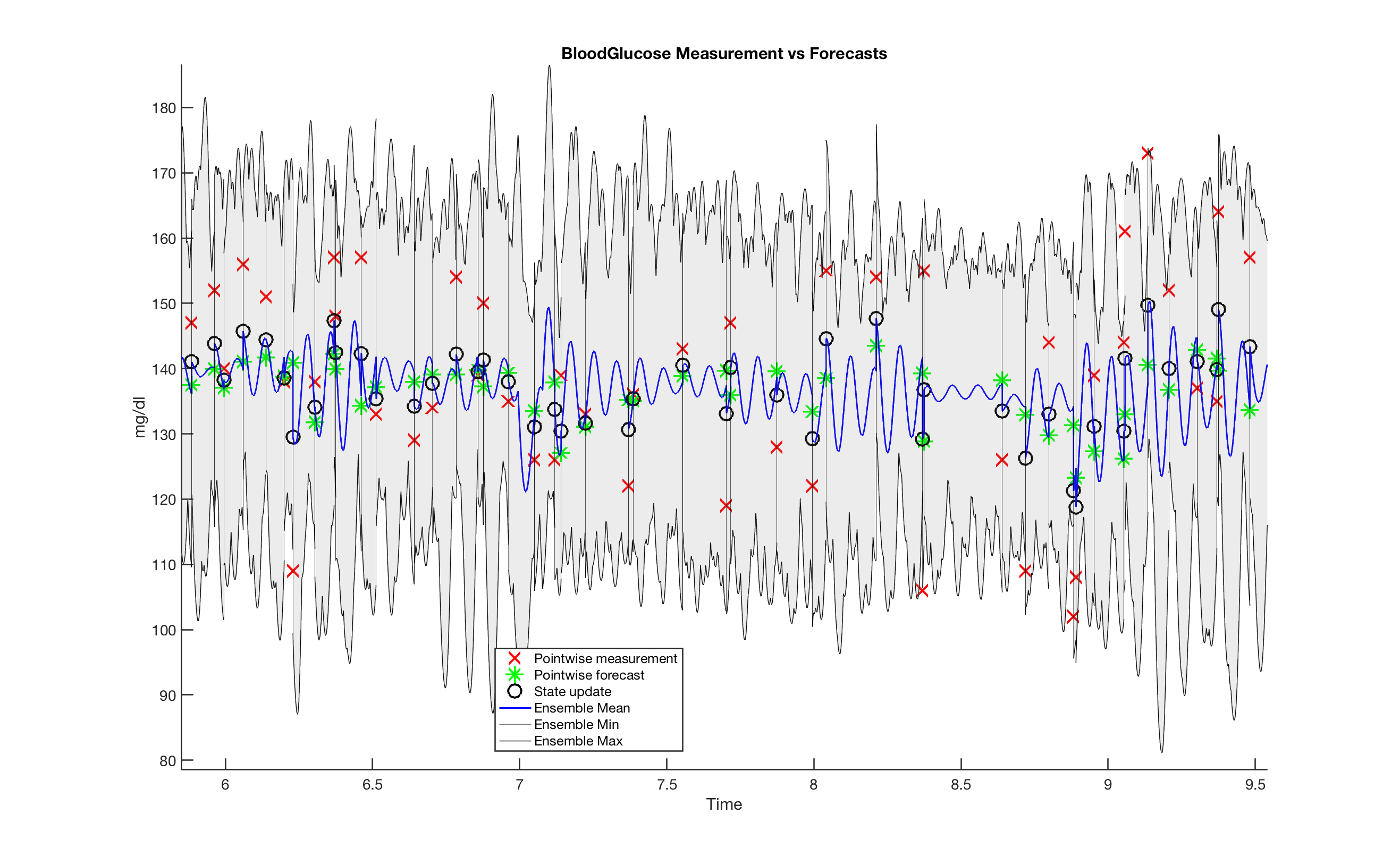}
\caption{This is an example of accurate state forecasting; these plots demonstrate the UM-EnKF $50$-particle ensemble estimating the $P_H$ parameter set forecasting patient 426. Notice that the forecast appears to begin to accurately track the patient after about 1-1.5 days and the future data points generally lie within the $50$-particle forecast ensemble, meaning that the data lie within the forecast ensemble uncertainty.}
\label{fig:426umh2p50_2} 
\end{figure*}
\begin{figure*}
\begin{subfigure}{0.6\textwidth}
        \centering
	\includegraphics[scale=0.11]{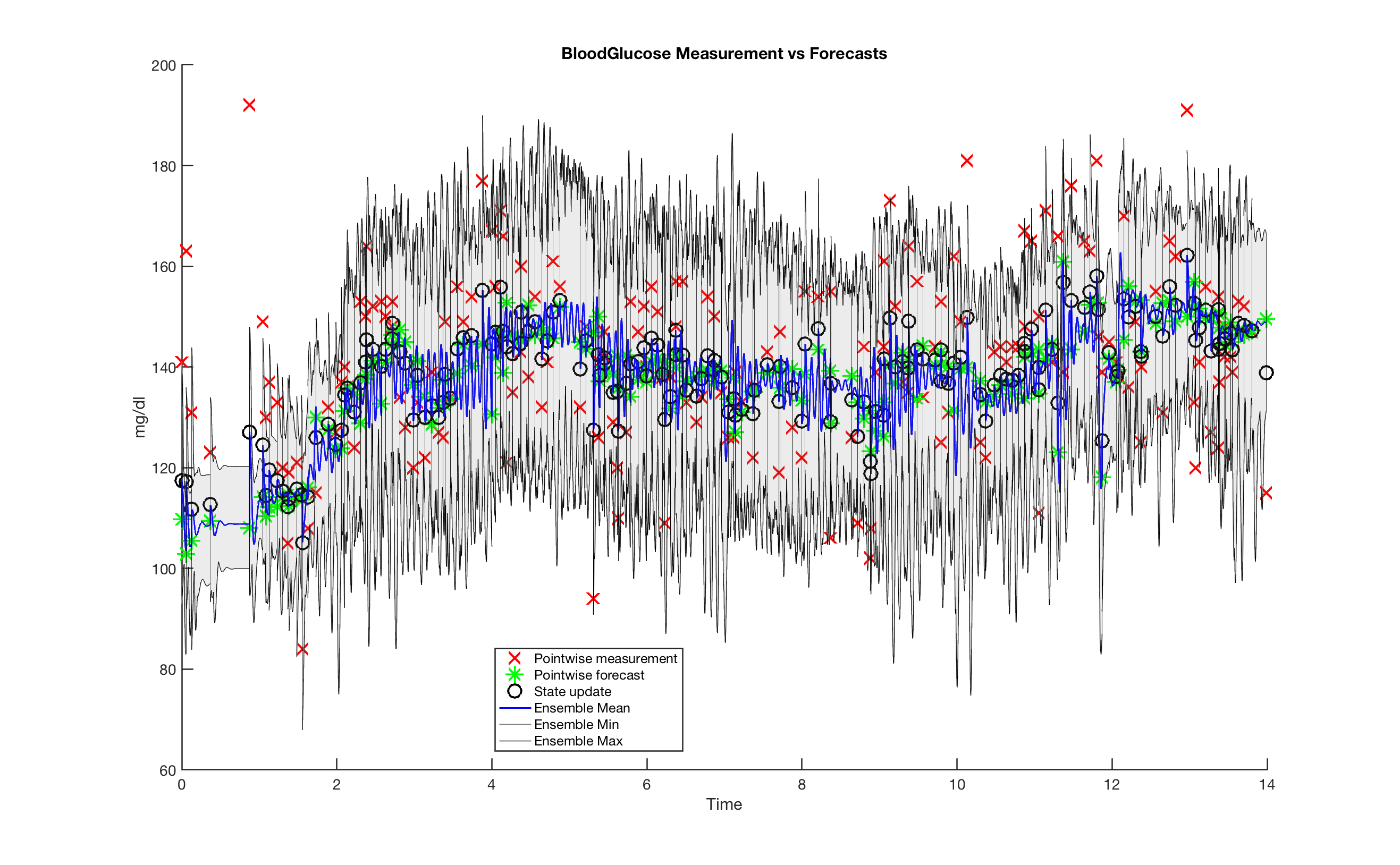}
\end{subfigure}
\begin{subfigure}{0.4\textwidth}
        \centering
\includegraphics[scale=0.21]{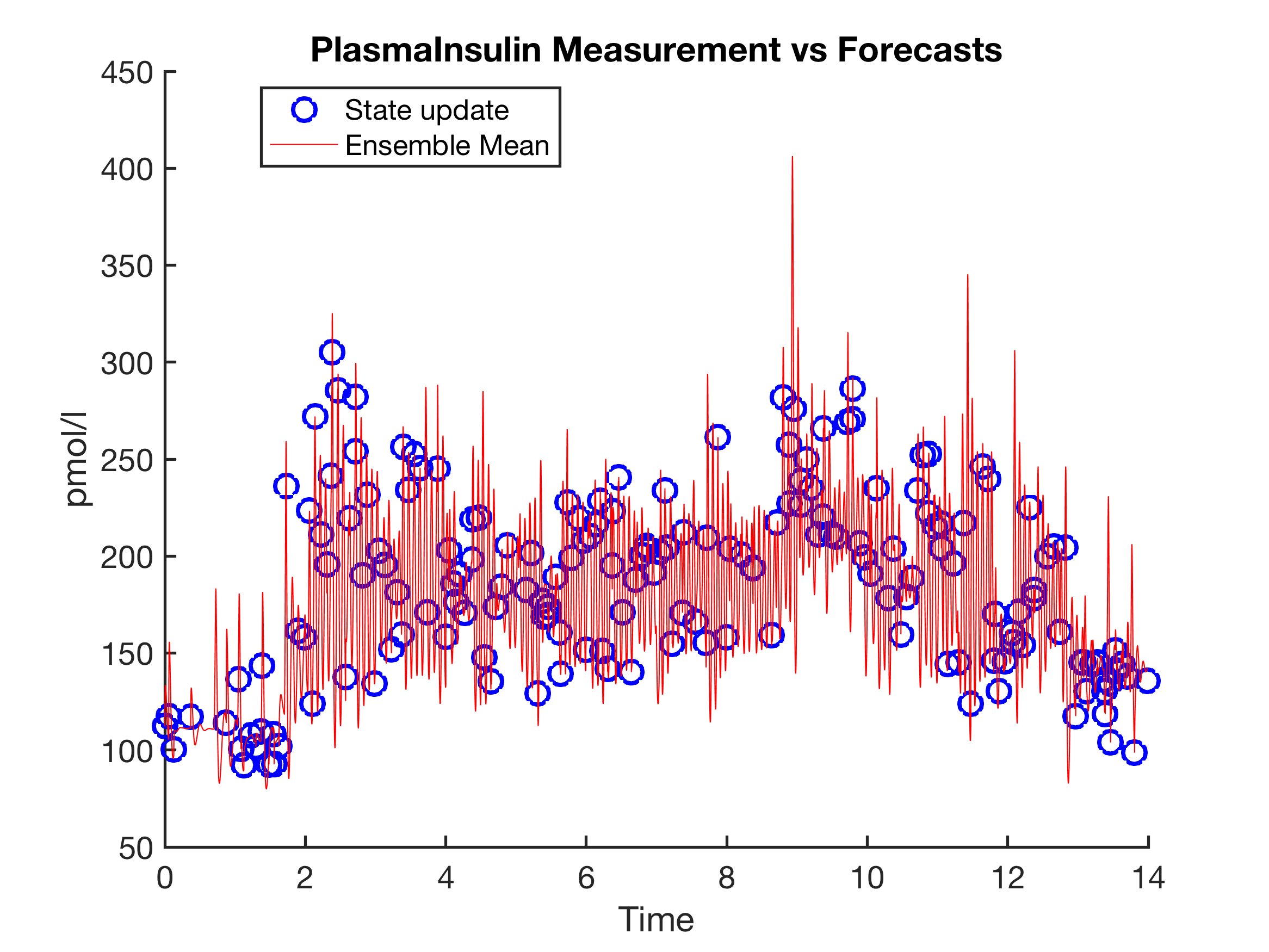} \\
\includegraphics[scale=0.21]{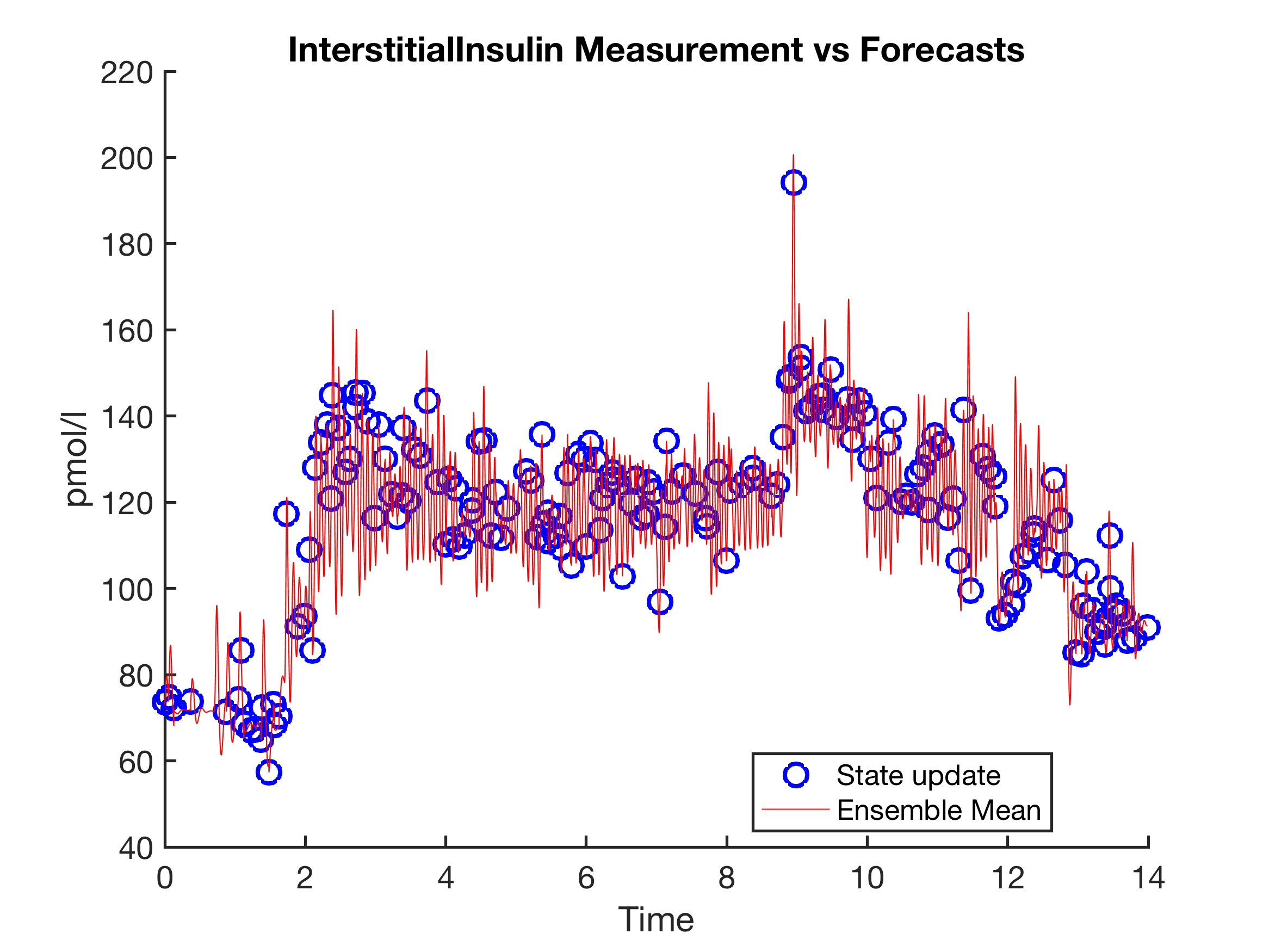}
\end{subfigure}
\begin{subfigure}{1\textwidth}
        \centering
\includegraphics[scale=0.3]{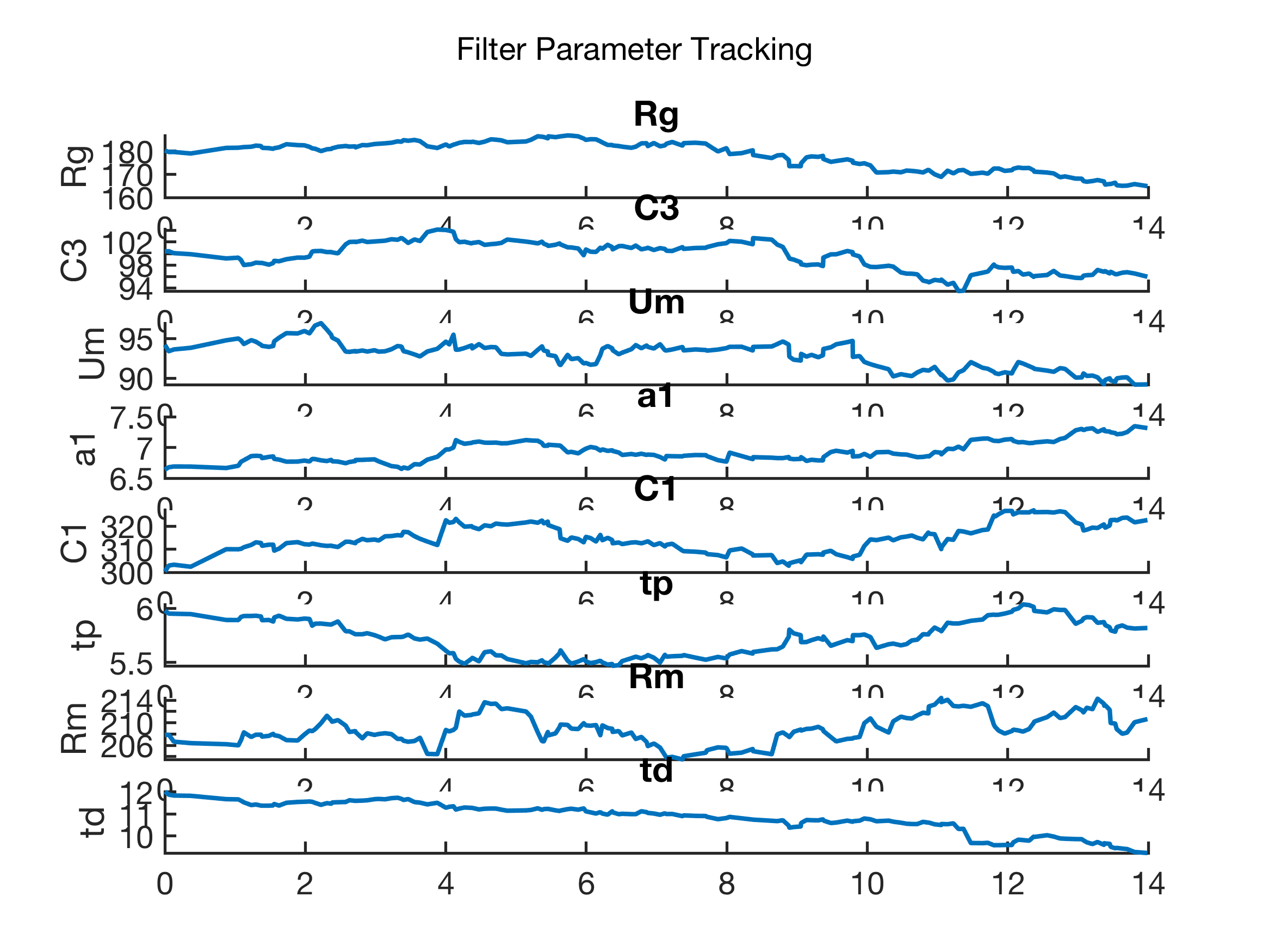}
\end{subfigure}
\begin{subfigure}{1\textwidth}
        \centering
\includegraphics[scale=0.27]{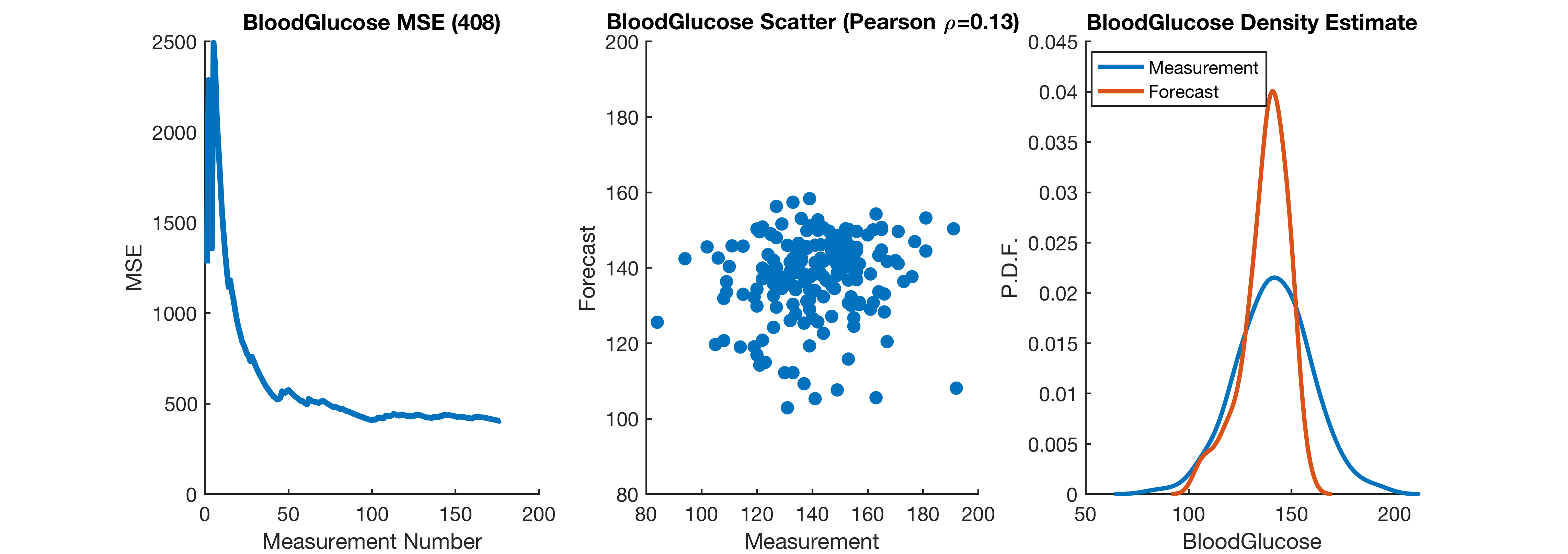}
\end{subfigure}
\caption{A detailed evaluation of the UM-EnKF estimating $P_H$-parameters forecasting patient 426 well and quickly. The lower panel (MSE) shows the forecast error is only slightly above the measurement error, while the ensemble forecast mean tends to underestimate the upper tails of the glucose distribution and point-wise correlation low. The middle panel shows all eight \emph{Houlihan} parameters varying over time as the patient evolves---parameters are evolved at every data point as they are inferred and are otherwise constant. The unmeasured insulin---\emph{note there are no insulin measurements, only simulated states and state updates}---shown in the upper panels, remains within reasonable ranges despite not being constrained by measurements or any other means, a result that will contrast with other later cases. And the ensemble forecast mean, detailed in the upper panel, tracks the non-stationarity of the patient well over time; it is the distribution of the forecast ensemble mean that is used for the distributional evaluation in the lower right panel.}
\label{fig:426umh2p50_1} 
\end{figure*}

\paragraph{\textbf{Example of effective EnKF-based forecasting}} Figures  \ref{fig:426umh2p50_2}-\ref{fig:426umh2p50_1} show an example the UM-EnKF estimating $P_H$-parameters and states for a patient with an MSE near 300---approaching the magnitude of measurement error---within a day. Figure \ref{fig:426umh2p50_2} shows two regions of glycemic dynamics characterized by $50$ particles of the UM-EnKF forecast ensemble, forecast mean, and measurements. The first plot demonstrates the path to convergence---the UM-EnKF has converged by day 1.5---and the second plot shows converged forecasting. The ensemble captures the uncertainty in the forecast well: with $50$ particles the data lie within the ensemble forecast. The forecast mean glucose density in Fig. \ref{fig:426umh2p50_1} shows the UM-EnKF captured the lower distribution boundary more accurately than the upper boundary. Correctly predicting glycemic tails is canonically challenging with sparse data. Nevertheless, if these events are dangerous for a patient, then they remain important to forecast correctly. Figure \ref{fig:426umh2p50_1} shows that while overall MSE error magnitude is low, the correlation between ensemble mean forecast and measurement is low but is significant. The low point-wise correlation is due to difficulties resolving the glucose oscillation frequency with sparse data and the impacts of dyssynchrony between the model and the data, cf Fig. 2 of \cite{jamia_da}, issues that will become more significant in later sections. The UM-EnKF, when estimating $P_H$-parameters and states, is able to represent the glycemic oscillations but the forecast mean is only weakly correlated with the small fluctuations. Finally, Fig.\ref{fig:426umh2p50_1} shows the glucose forecast ensemble mean and the parameter evolution track the non-stationarity of the patient, and the unmeasured insulin dynamics remain with plausible ranges.

\begin{table*}
\centering
\resizebox{\textwidth}{!}{
{
\begin{tabular}{|l| |l| |l|  }
  \hline
  \multicolumn{3}{|c|}{ EnKF MSE after 24 data points ($<24$ hr) including tube feed and insulin} \\
  \hline \hline
  id & $UM(P_H = R_g, C_3, U_m, a_1, C_1, t_p, R_m, t_d)$ & $UM(P_{RH}=R_g, a_1, C_1, C_3, t_d)$  \\ \hline \hline
  296 & 1254&2963\\ \hline
426 & 321&311 \\ \hline
430 & 2654&16785 \\ \hline
456 & 444&413  \\ \hline
489 & 9512&10466\\ \hline
585 & 730&2795 \\ \hline
593 & 5014&19883 \\ \hline
646 & 2083&4702 \\ \hline
851 & 4799&13792 \\ \hline   \hline \hline
   \multicolumn{3}{|c|}{ EnKF MSE after 24 data points ($<24$ hr), adding incidental IV glucose infusion but not bolus} \\
  \hline \hline
296 & 1890 & 2729    \\ \hline
 426 &  356&  370  \\ \hline
 430 & 3440& 16970  \\ \hline
456 &   479&503  \\ \hline
 489 & 9497&9716  \\ \hline
 585 & 891&3576   \\ \hline
 593 & 5538&19951 \\ \hline
 646 & 1889&5153   \\ \hline
 851 &  9792&18425  \\ \hline \hline
 \multicolumn{3}{|c|}{ EnKF MSE after 24 data points ($<24$ hr), adding IV glucose bolus but no incidental IV glucose infusion} \\   \hline \hline
489 &  9480 &10462   \\ \hline
585 &  730 &2813  \\ \hline
 851 &  5651 &14195   \\ \hline
  \hline \hline

\end{tabular}
}
}
 \caption{Mean squared error (MSE) between forecasts and measurements for the UM-EnKF for both parameter configurations and for three different data configurations.}
   \label{table:simple_mse}
\end{table*}

\paragraph{\textbf{Forecasting accuracy across patients}} The mean squared error (MSE) for the UM-EnKF for all parameter sets and experiments are shown in Table \ref{table:simple_mse}. Not all patients have data of all types, and therefore are not included in all experimental frameworks. The results follow five themes.

\emph{\underline{Theme one:}} patient-specific factors influence inference success. Patients with less complex courses, e.g., without insulin intervention, are more accurately estimated with the UM-EnKF  and with fewer data points. This aligns with the failure point of other efforts \cite{29_clermont_pritchard2017modeling}.  Patients whose initial physiology is near nominal parameters at admission are estimated more accurately with fewer initial data points because the parameters do not require appreciable changes to achieve accurate forecasts. Online modeling is capable of adapting to non-stationary, but this capability has limitations and accurate parameter initialization will clearly help forecast accuracy.

\emph{\underline{Theme two:}} accurate representation of physiologic dynamics can impact forecast mean and accuracy. Even when the UM-EnKF has an MSE approaching measurement error, the ensemble mean is typically weakly correlated with the data. Estimating glycemic oscillations---minimally their existence and amplitude---helps estimating glycemic variance. However, exact matching of these oscillations may or may not be necessary for clinical decision-making. Additionally, all models approximated with unconstrained UM-EnKF can fail to capture measurements from the tail of their distribution. It is essential to consider the tolerability of these types of forecasting errors with respect to clinical decision-making. Finally, trends in the parameter tracking could be potentially useful for predicting the patient disease trajectory.

\emph{\underline{Theme three:}} canonical skill scores and evaluation are not always meaningful in this context. Neither MSE nor correlation represent quantities directly relevant for clinical decision-making because they do not reflect the impact of forecasts on decision-making. Additionally, there is often discord between skill scores without a productive method for selecting the most useful skill score. For example, as we will see later the MSE can be quite high while the glucose forecast and measurement distributions are quite similar, leading to discord among evaluation metrics.

\emph{\underline{Theme four:}} including incidental IV glucose drip does not reduce, and can elevate, forecast error.  This result was expected because the measurement of non-bolus IV glucose is highly error prone, e.g., times of administration can be off by 10s to 100s of minutes, and typically represents a small amount of ingested carbohydrates.

\emph{\underline{Theme five:}} including IV glycemic bolus does not negatively impact forecast error. These bolus events are relatively rare and are followed by an increase in measurement frequency from measurements every $1-2$ hours to measurements every $15$ minutes allowing the filter to be much more synchronized to the patient. This is an important result because glycemic boluses are highly clinically relevant and represent adverse events.

\subsection{Limiting factors of unconstrained inference}
Table \ref{table:simple_mse} demonstrated that unconstrained UM-EnKF is not accurate for all patients. While model error surely contributes to the forecast error, here we will focus on two serious inference problems that limit the use of unconstrained inference for models such as the UM with sparse data.

\begin{figure*}
\begin{subfigure}{0.6\textwidth}
        \centering
	\includegraphics[scale=0.5]{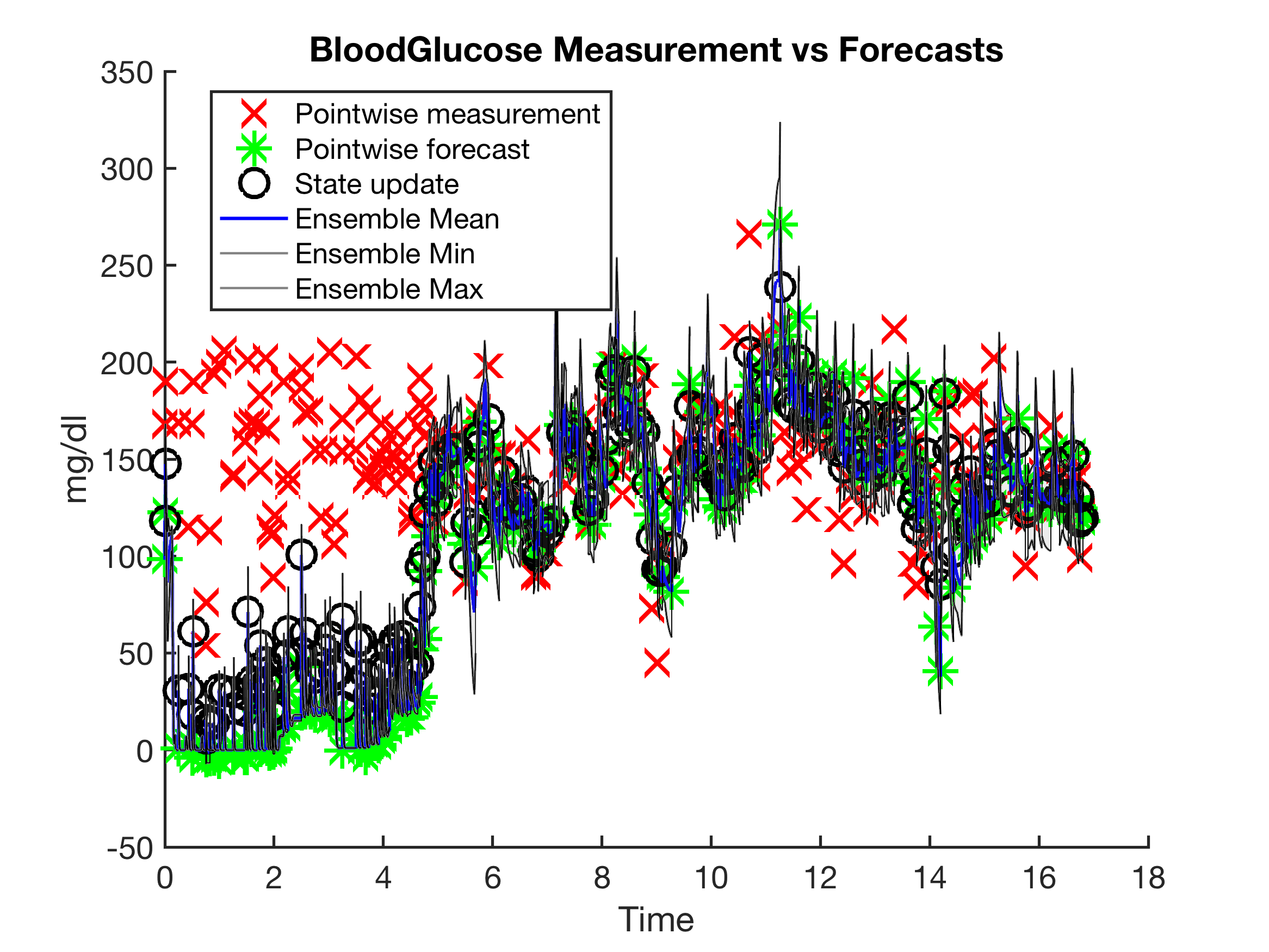}
\end{subfigure}
\begin{subfigure}{0.4\textwidth}
        \centering
\includegraphics[scale=0.25]{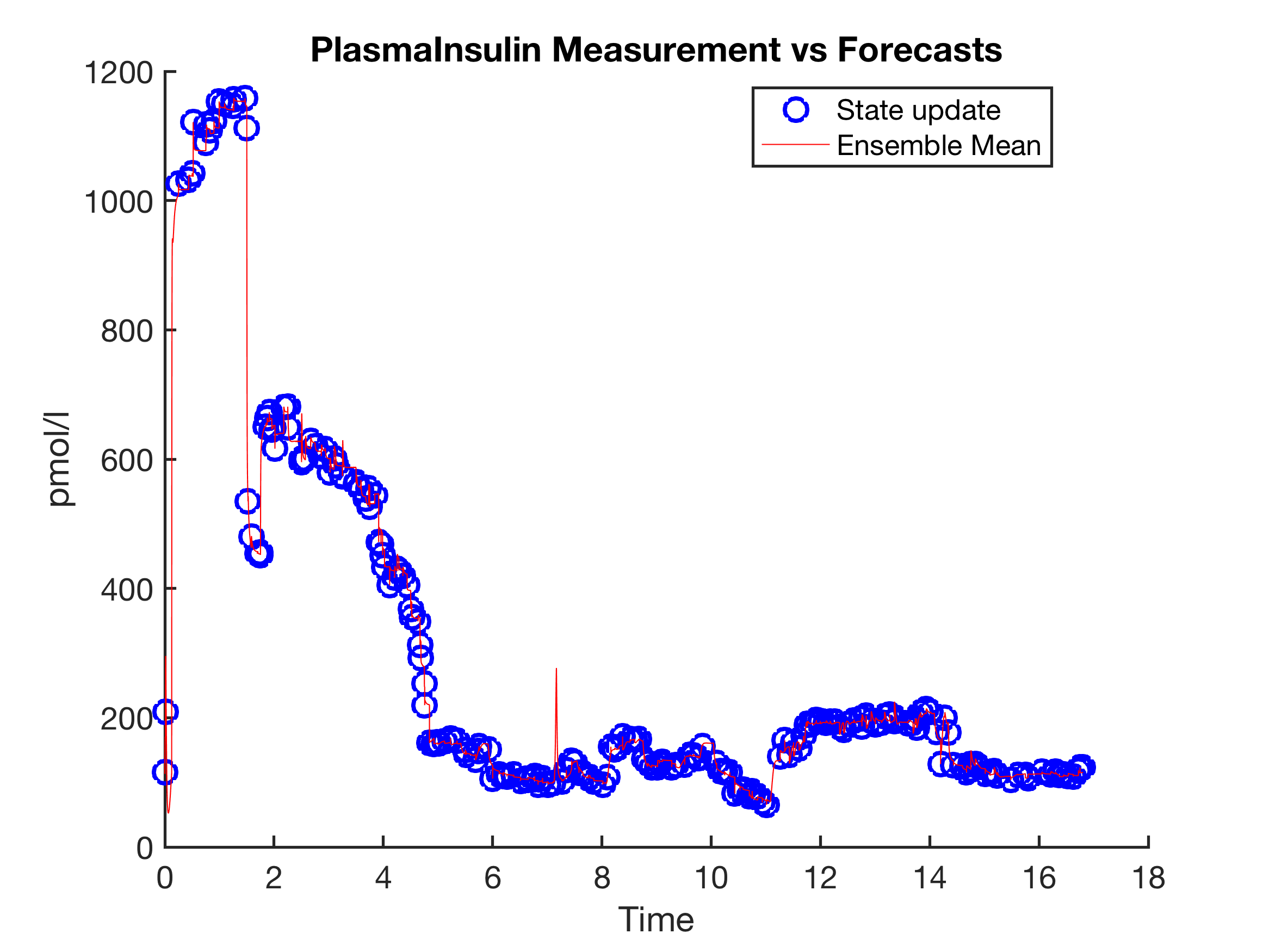} \\
\includegraphics[scale=0.25]{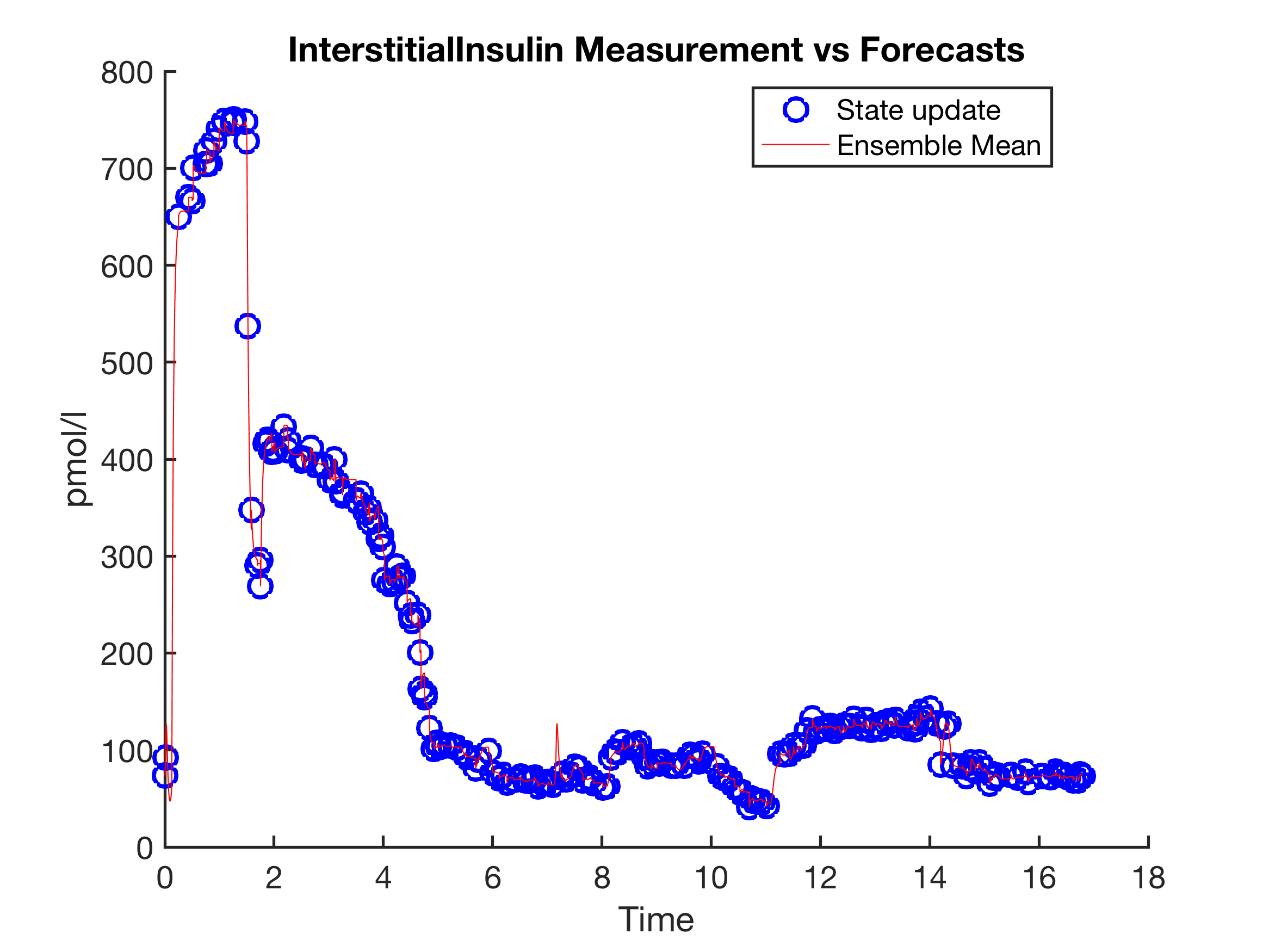}
\end{subfigure}
\begin{subfigure}{0.95\textwidth}
        \centering
\includegraphics[scale=0.3]{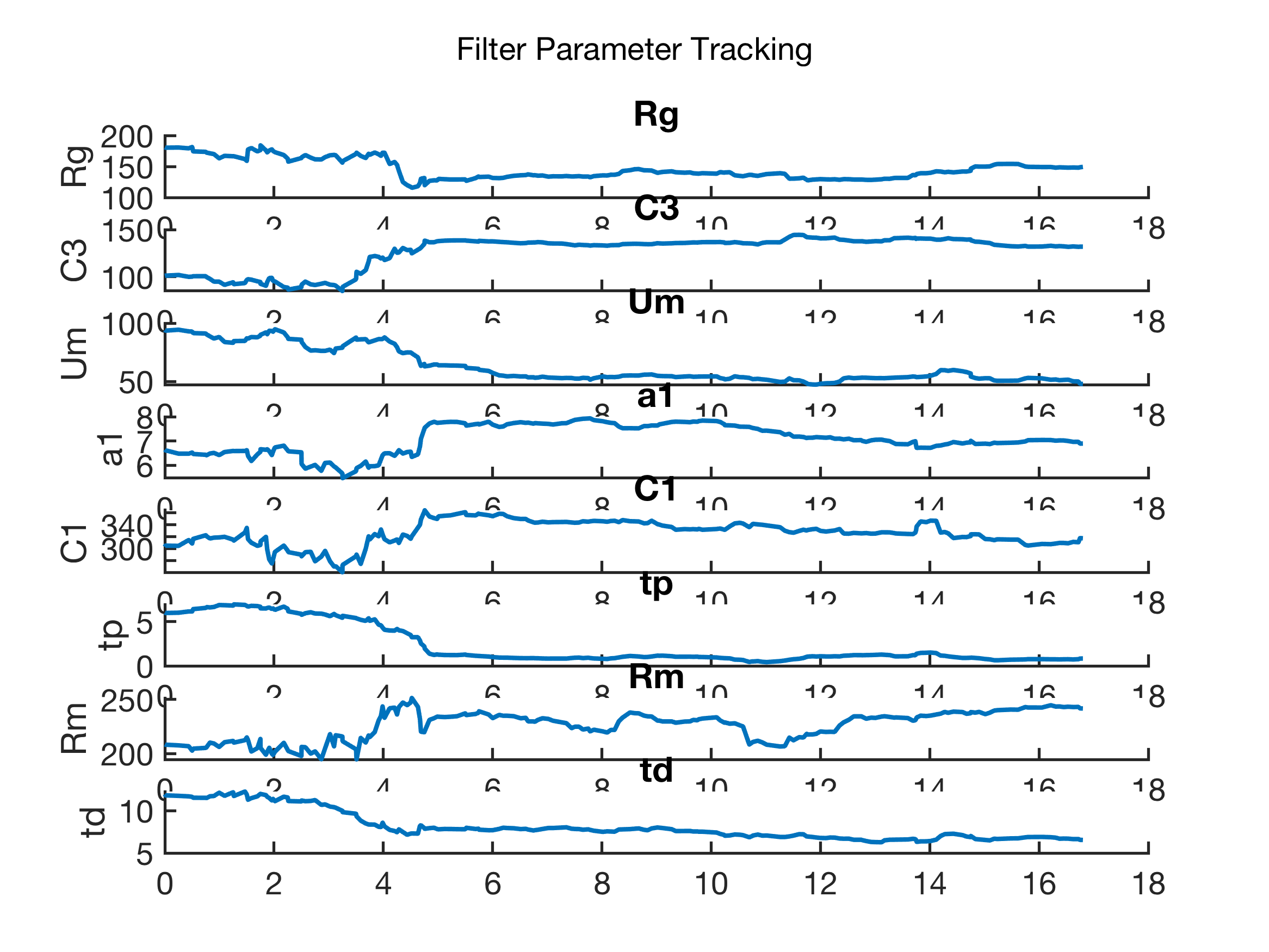}
\end{subfigure}
\begin{subfigure}{0.95\textwidth}
        \centering
\includegraphics[scale=0.28]{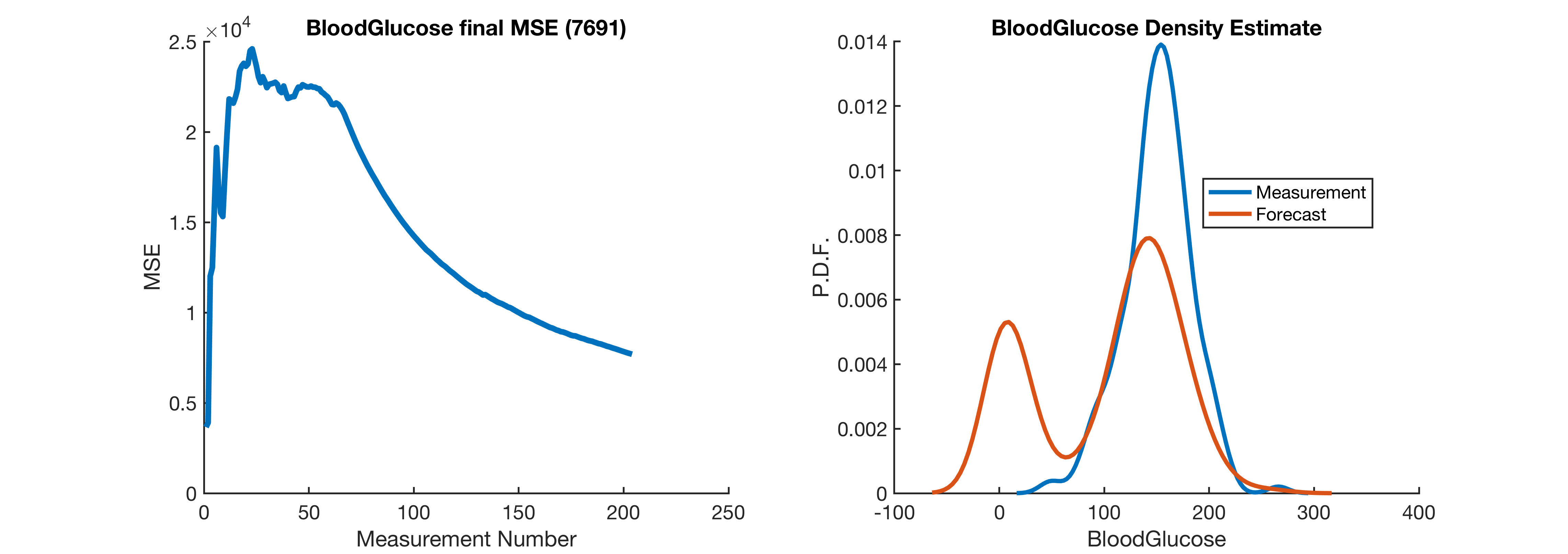}
\end{subfigure}

\caption{This figure, where the UM-EnKF is estimating patient 593 with the $P_H$ parameter set, demonstrates forecast failure due to the lack of identifiability of insulin. The upper right panels shows insulin dynamics---there are no insulin measurements to constrain the inference---that clearly lie outside a sensible range in the first $4-4.5$ days of the patient's stay.  The evaluation metrics in the bottom panels show the multiple glucose equilibria, one $\sim 0$ leading a very high MSE. The upper left panel shows the the ensemble forecast mean that lies around $0$ until around day $4-4.5$ at which point the model begins to track the patient accurately.  The parameter estimates in the middle panel show the point where the model finds a good solution in parameter space, around day $4.5$, from which point the parameters move smoothly and track the patient non-stationarity over time. Despite the identifiability problems, the model is eventually able to entrain to the patient because of the continuous parameter adaptation of the UM-EnKF; the UM-EnKF is not always so lucky and for some patients is not ever able to attain accurate entrainment when initialized with the nominal parameters.}
\label{fig:593umh2p15}
\end{figure*}

\paragraph{\textbf{Model identifiability failure}} Model identifiability may be defined as the ability to uniquely solve for parameters and then predict states, given available data. The models we use are nonlinear, making analytic proof of identifiability often impossible, but given the available data we know the models are not identifiable. The ultradian model has three states, one measured, BG, and two unmeasured, plasma and interstitial insulin, along with $21$ unmeasured parameters. Because of this, it is not identifiable given realistic data, regardless of the structural identifiability properties of the ultradian model. The Houlihan approach uses machine learning to reduce this issue by limiting the parameter space inferred, but this method does not eliminate identifiability failure because the insulin states are never measured. One pathway to resolving this problem is to develop models without insulin as an explicit state variable \cite{melike2019simple}. Nevertheless, even models that exclude insulin as a state variable can have identifiability problems because clinical measurement patterns can make estimating glycemic response non-unique \cite{melike2019simple}.  Meaning, for robust inference with clinical data, we will always have to manage inference with missing measurements and non-identifiable models. But the case here reveals the inference problem clearly: because glucose and insulin are coupled, and because the plasma and interstitial insulin levels are free states unconstrained by data, it is natural for the EnKF, or any other inference method, to adjust insulin levels widely. When this freedom is exercised, it increases the time required for the model estimate patient and can cause catastrophic estimation failure. Figure \ref{fig:593umh2p15} demonstrates this problem, the UM-EnKF initially forecasts glucose $\sim 0$ and it takes the UM-EnKF $4.5$ days of data to find model parameters that represent the patient accurately. For more complex cases, and for more restricted models, the UM-EnKF is not able to synchronize the model with the data.

\paragraph{\textbf{Accurate representation of glycemic dynamics}} The goal of clinical management of BG is to keep patients within a pre-specified range. Therefore, one of the key needs for BG forecasting is an accurate estimate of the variance of BG dynamics. Continuous tube feeding induces glycemic oscillations \cite{sturis_91}. The primary source of the oscillations are delayed responses, e.g., of glucose production by the liver \cite{sturis_91}, with adipose and muscle tissue, \cite{artie_muscle_adipose}, although there may also be additional complex dynamical issues \cite{karamched2020delayinduced}. Consequently, making model forecasts potentially useful for decision making likely requires resolving BG oscillations to some degree of accuracy. Models fail to elicit oscillations for three reasons. First, the functional form of the model can fail to have glycemic oscillations. Second, a model capable of producing glycemic oscillations can be restricted such that it is unable to produce oscillation, e.g., by fixing parameters that control oscillatory dynamics into ranges where the model does not elicit oscillations. The highly restrictive Houlihan-selected parameters tend to make the ultradian model too inflexible to represent BG oscillations. Third, the inference method can calculate optimal parameters that extinguish oscillations. Often in the sparse data setting, one local minima is to have the model represent the constant mean of the data. Given sparse enough data, the Kalman filter machinery, which in its linear from estimates the conditional mean, naturally identifies solutions that may minimize oscillations, a topic we will take up in the discussion.

\subsection{Improving the forecast with knowledge using constrained EnKF}

Lying at the heart of the failure of the EnKF to estimate a patient quickly and accurately is data sparsity: given data and the complex physiology, we must make the model space as simple as possible but no simpler. We restrict the models in two ways. First, we use the Houlihan method \cite{houlihan} to select a minimal subset of parameters to estimate. Second, we use a constrained EnKF (CEnKF) \cite{cenkf}. Constraining inference addresses two problems: (i) constraints reduce identifiability issues, e.g., with the lack of insulin measurement, and (ii) constraints disallow impossible physiology.

\begin{figure*}
\begin{subfigure}{0.6\textwidth}
        \centering
	\includegraphics[scale=0.12]{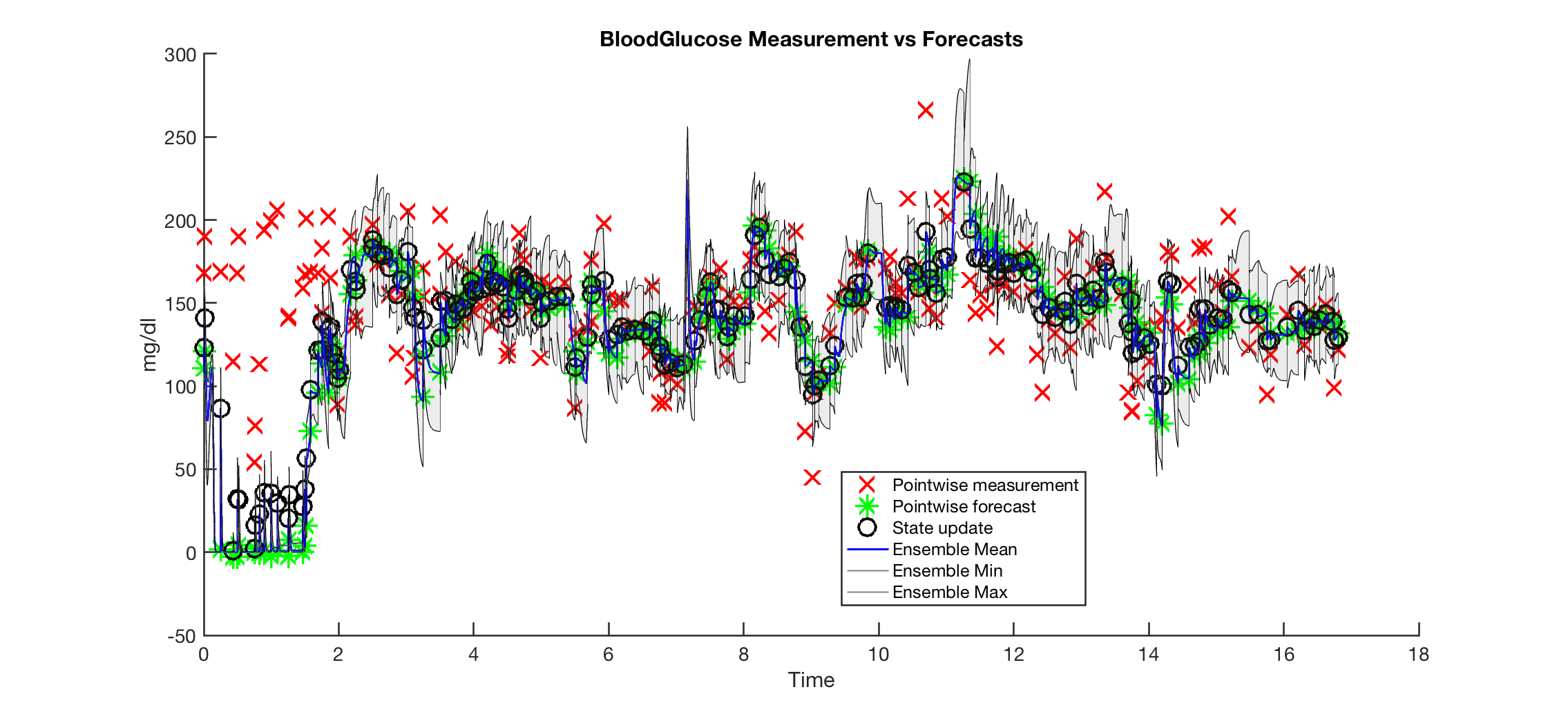}
	\includegraphics[scale=0.0735]{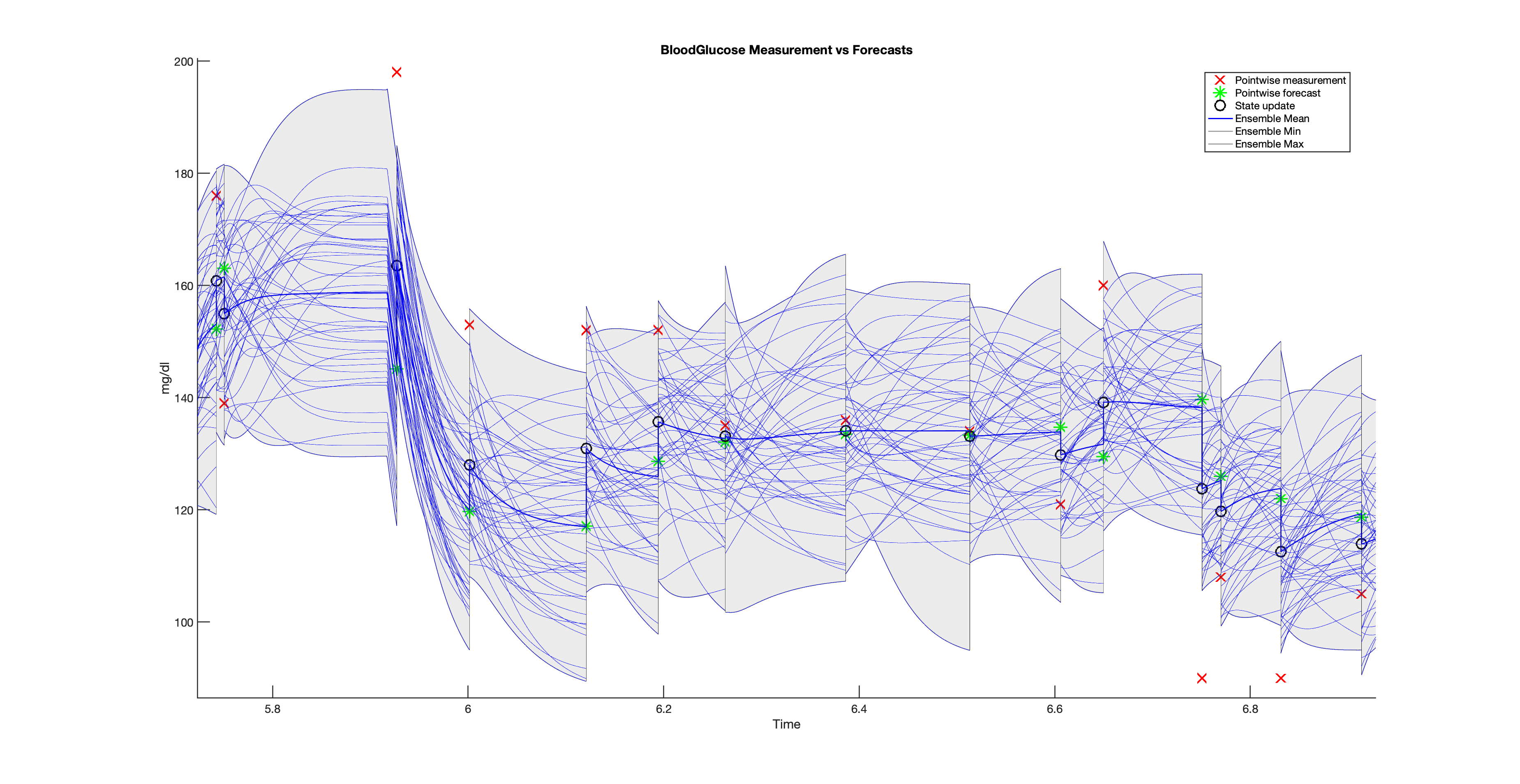}
\end{subfigure}
\begin{subfigure}{0.4\textwidth}
        \centering
\includegraphics[scale=0.25]{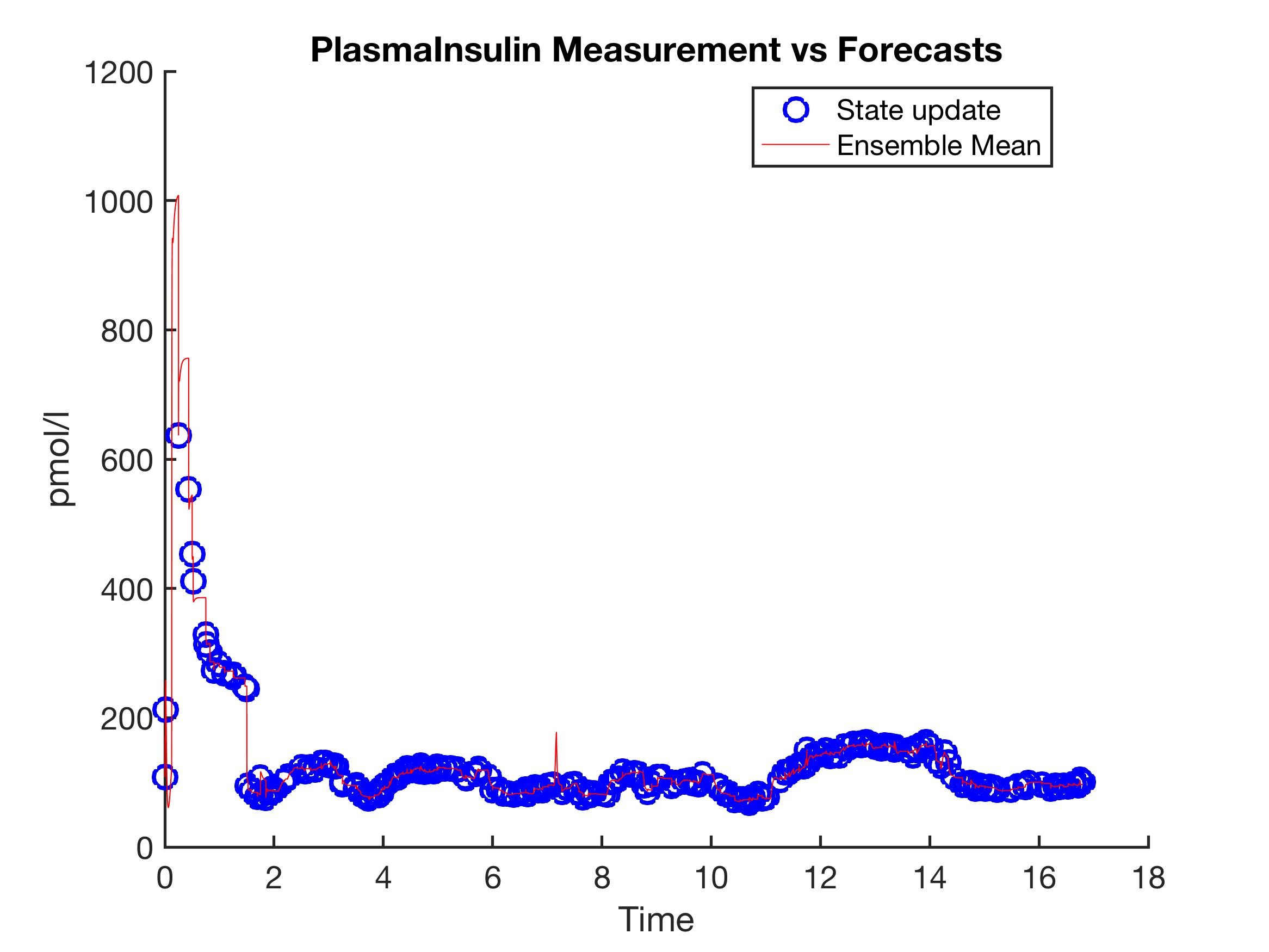}
\includegraphics[scale=0.25]{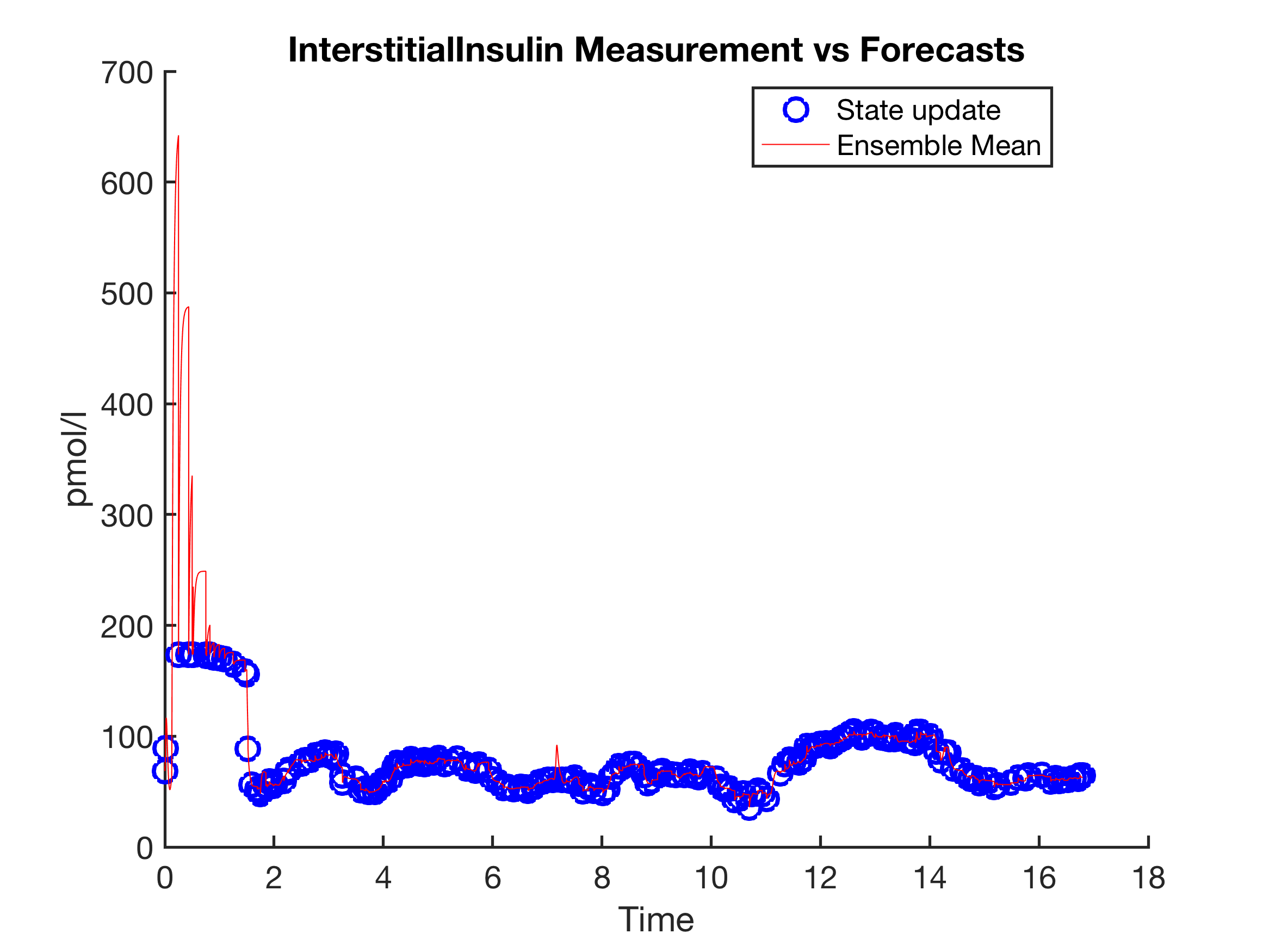}
\end{subfigure}
\caption{This figure demonstrates how constraints--severely constrained insulin (IS)---improve forecast accuracy and robustness applied with the UM-CEnKF, compared directly with the UM-EnKF forecast results shown in Fig. \ref{fig:593umh2p15}. The right panels show insulin dynamics are quickly constrained to lie within the realistic constraint boundaries, resulting in the model entraining to the patient within $1-1.5$ days instead of $4-4.5$ days. The upper left panel shows the forecast ensemble converging and estimating the patient's mean BG; the variance in BG is under-estimated. We hypothesize sparse measurements cause inaccurate representation of the dynamics or the invariant measure, shown in the lower left plot, causing under-estimation of the variance. Our claim, based on the lower left plot where showing ensemble trajectories, is that the UM-CEnKF is not able to resolve and solve for the parameters that lead to the correct oscillatory frequencies present in the real system due to data sparsity.}
\label{fig:593umh2p50_c_is}
\end{figure*}

\begin{figure*}
\begin{subfigure}{0.95\textwidth}
        \centering
	\includegraphics[scale=0.4]{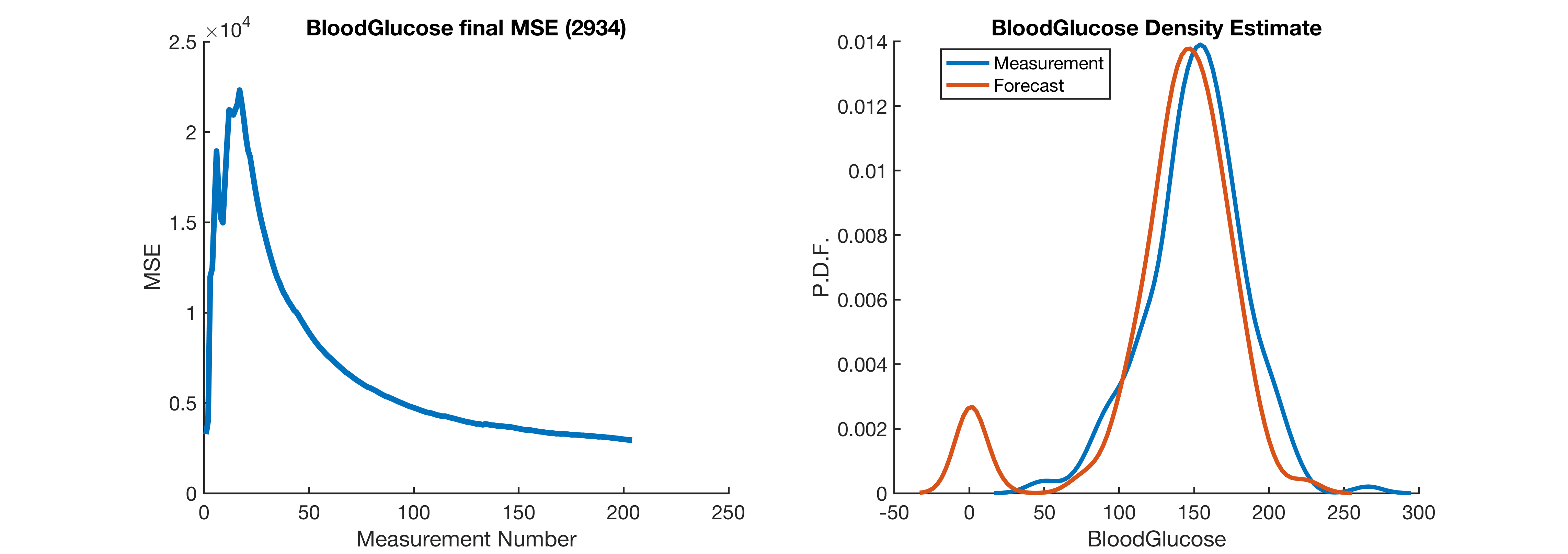}
\end{subfigure}
\begin{subfigure}{0.5\textwidth}
        \centering
\includegraphics[scale=0.375]{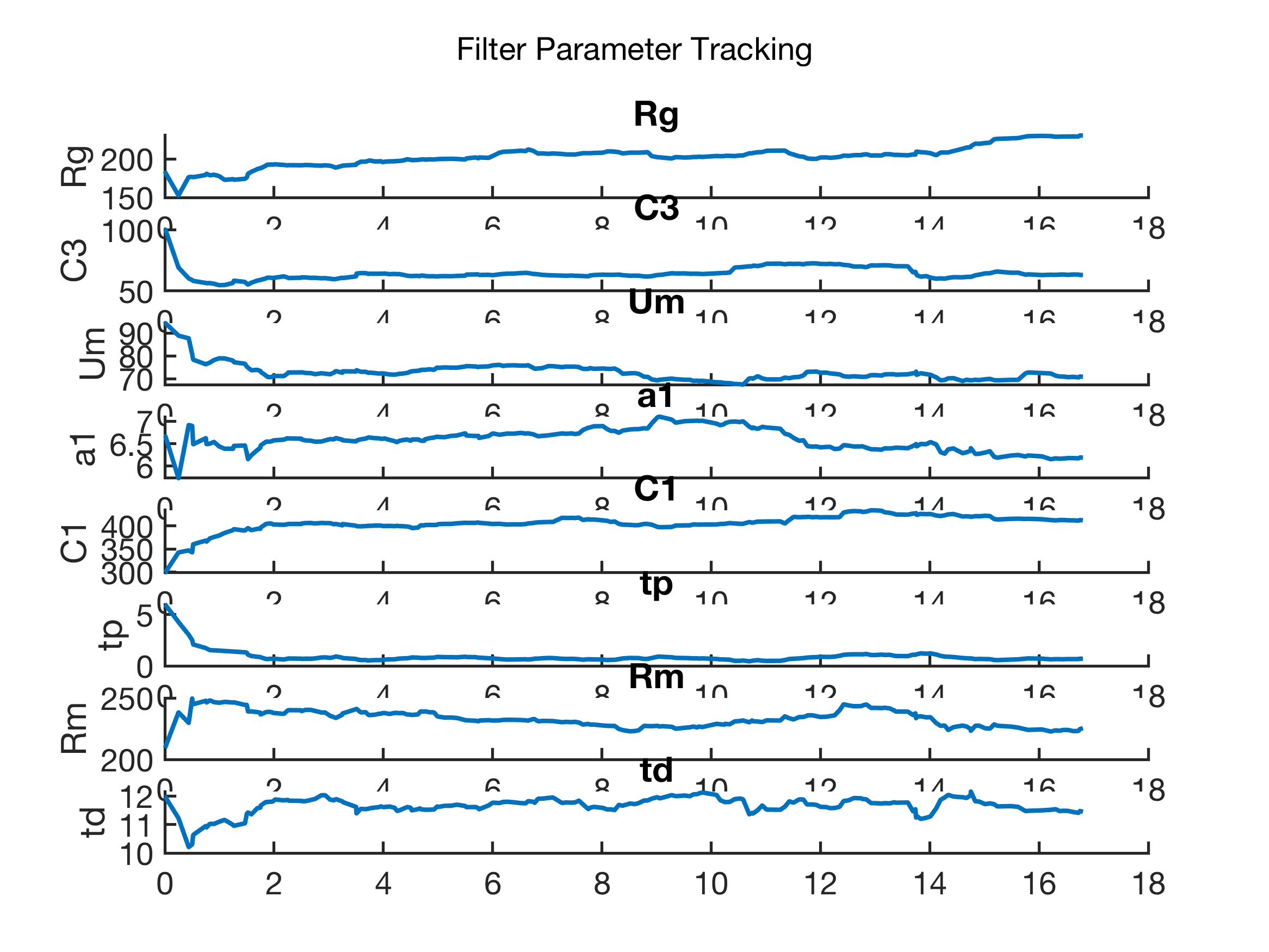}
\end{subfigure}
\begin{subfigure}{0.5\textwidth}
        \centering
\includegraphics[scale=0.375]{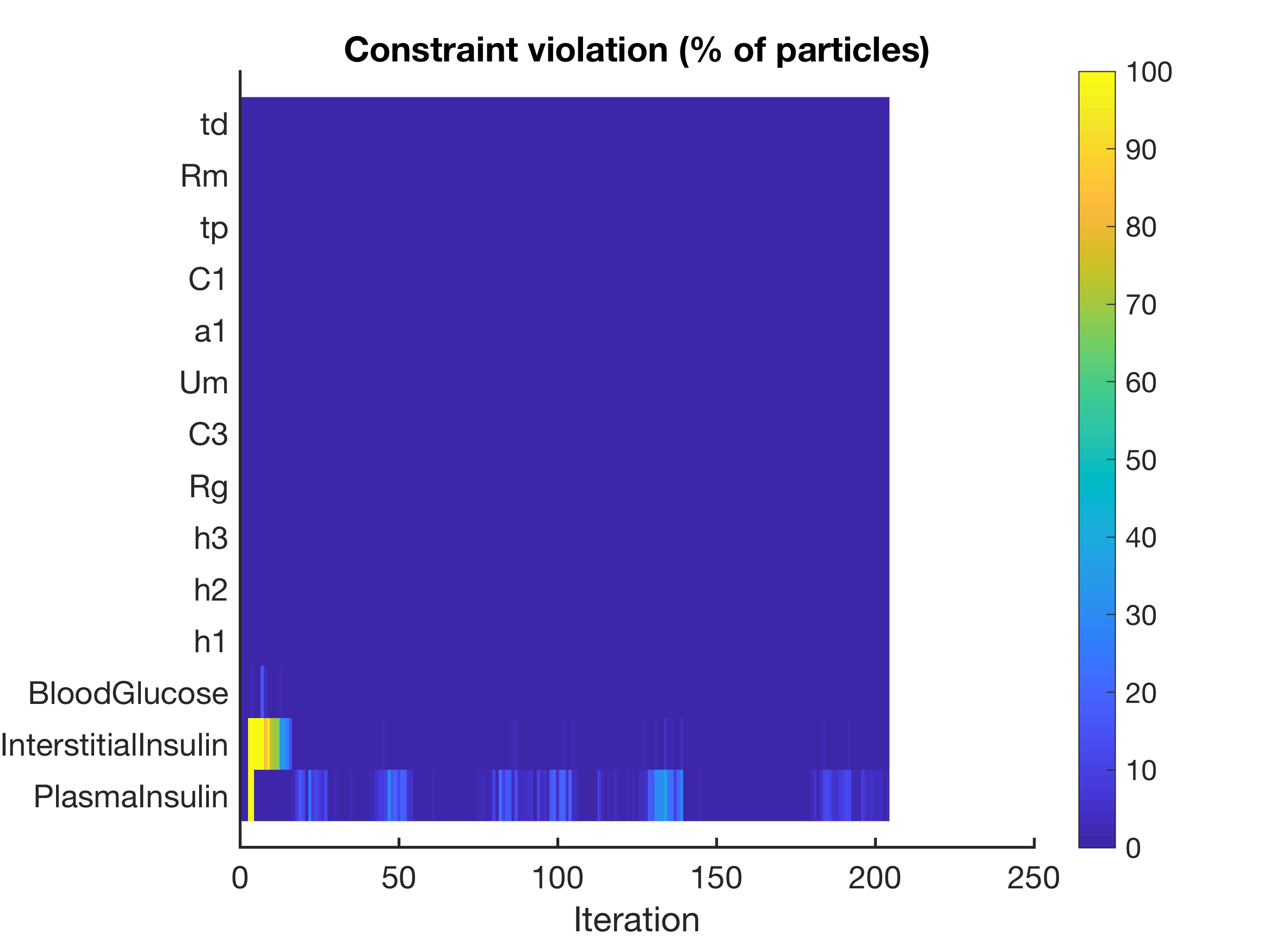}
\end{subfigure}
\caption{This figure details evaluation metrics for the UM-CEnKF for patient 593. The upper right plot shows the multiple glucose equilibria, but the bad solution, BG$\sim0$, is eliminated within a day resulting in a MSE reduced by a factor of between $5$ and \textbf{50}, shown in the upper left plot. The BG ensemble mean forecast distribution, showing in the upper right panel, approximates the measured distribution nearly perfectly after the UM-CEnKF entrains to the patient. The parameter estimates, shown in the lower left panel, show the time-point where the model finds a good solution in parameter space around day $1-1.5$, from which point the parameters change smoothly and track the patient non-stationarity over time. Finally, constraints violation of the unmeasured insulin is restricted to the first 20 or so data points, or within the first day, after which the UM-CEnKF  finds model parameters that allow for accurate BG forecasting. Overall, the constraints reduce identifiability failure due to unmeasured insulin and allow for accurate forecasting with $\sim 25 \%$ of the data $20$ data points.}
\label{fig:593umh2p50_c_is_eval}
\end{figure*}

Figure \ref{fig:593umh2p50_c_is} shows the UM-CEnKF producing more accurate BG forecasts with less data. The UM-CEnKF constrains insulin between $75-275$, clinically reasonable values, and contrasts well with the unconstrained UM-EnKF shown in Fig. \ref{fig:593umh2p15}. The UM-CEnKF enables accurate forecasts after $1-1.5$ days compared to $4.5-5$, decreases MSE by a factor of $5$, and the BG forecast distribution, shown in Fig.  \ref{fig:593umh2p50_c_is} as forecast uncertainty and in Fig. \ref{fig:593umh2p50_c_is_eval} as a distribution over all time points, is approximated nearly perfectly including the distribution tails. Figure \ref{fig:593umh2p50_c_is}, shows that insulin constraints are violated at the beginning of the ICU stay, after which, the model tracks the patient. This example showcases knowledge-driven constraints balancing model flexibility and fidelity, allowing the UM-CEnKF to forecast given sparse data.

\paragraph{\textbf{Constrained EnKF experimental framework}} To evaluate the effectiveness of UM-CEnKF we ran a series of computational experiments whose labels and the corresponding constraints on parameters are shown in Table \ref{table:constraints}. The impacts of different constraints on forecast accuracy is detailed in two analyses shown in Table \ref{table:constraint_mse_ratio}. First, for a given patient we report the ratio of MSE of the UM-CEnKF over MSE of the UM-EnKF, denoted $\omega$. When $\omega >1$ the constraints increase forecast error, when $\omega =1$ the constraints do not affect the forecast, and when $\omega <1$ the constraints reduce forecast error. Accordingly, we split the effect of constraints into four categories: harmful, $\omega \geq 1.1$, no harm, $0.9<\omega<1.1$, improvement, $0.6<\omega \leq 0.9$, and substantial improvement,  $\omega \leq0.6$. Then we report the fraction of patients whose forecasts fell into the respective categories.

\begin{table*}
\begin{tabular}{|l| |l| |l|}
  \hline
  \multicolumn{3}{|c|}{Ultradian constraint experiment structure} \\
  \hline \hline
Exp. No. & constraints & Abbreviation \\ \hline
1 & glucose mild & gm  \\ \hline
2 & glucose severe & gs \\ \hline
3 & insulin mild & im \\ \hline
4 & insulin severe & is \\ \hline
5 & insulin ultra severe & ius \\ \hline
6 & glucose and insulin mild & gim \\ \hline
7 & glucose and insulin severe & gis \\ \hline
8 & glucose and insulin and parameters mild & gipm \\ \hline
9 & glucose and insulin and parameters severe & gips \\ \hline
10 & glucose and parameters mild, insulin ultra severe & gpmius  \\ \hline
11 & glucose and parameters severe, insulin ultra severe & gpsius \\ \hline
 \hline \hline
\end{tabular}

\begin{tabular}{|l| |l| |l| |l|}
  \hline
  \multicolumn{4}{|c|}{Ultradian constraints} \\
  \hline \hline
state/parameter & mild & severe & ultra severe \\ \hline
glucose & (20, 1000)  & (40, 400)  & --- \\ \hline
plasma insulin & (10, 400) & (75, 275) & (100, 250) \\ \hline
Interstitial insulin &(10, 400) & (75, 275) & (75, 175) \\ \hline
$t_p$  & (0.6, 60)  &(3,12)  & --- \\ \hline
$t_d$ & (1.2, 120) & (6, 24) & --- \\ \hline
$R_m$ & (20.9, 2090)  &  (104, 418) & --- \\ \hline
$a_1$ & (0.66, 66.7) & (3, 14) & --- \\ \hline
$C_1$ & (30, 3000) &  (150, 600) & --- \\ \hline
$C_3$ &  (10, 1000) & (50,200)  & --- \\ \hline
$U_m$ & (9.4, 940)  &  (47, 188) & --- \\ \hline
$R_g$ & (18, 1800) & (90, 360) & --- \\ \hline
\hline \hline
\end{tabular}
 \caption{The constrained EnKF (CEnKF) constraint experiments (top) and definitions and structure for the constraint experiments (bottom).}
   \label{table:constraints}
\end{table*}

\paragraph{\textbf{Constrained EnKF forecast impact across patients}} The UM-CEnKF results follow five themes shown in Figs \ref{fig:593umh2p50_c_is}-\ref{fig:593umh2p50_c_is_eval} and Table \ref{table:constraint_mse_ratio}.

\emph{\underline{Theme one:}} constraints on \emph{unobserved} states and parameters reduced identifiability failure due to unobserved states, decreasing the data required to track a patient while increasing forecast accuracy\emph{when the model as sufficient flexibility to well estimate the data.} Constraining unobserved states and parameters in lieu of measurements was the most powerful impact of constraints, and the most impactful constraints---the constraints that produced the largest gains in forecast accuracy and stability---were constraints applied to the unobserved states, specifically the plasma and interstitial insulin variables. When estimated using the Houlihan parameter set, $P_H$, about a third of the population has its MSE \emph{reduced by at least $40 \%$} when insulin is severely constrained, and the severe insulin constraint never reduces forecasting accuracy. Constraints had the opposite effect when the model was too restrictive, e.g., when estimated using the restrictive Houlihan parameter set, $P_{RH}$, discussed in theme two.

\emph{\underline{Theme two:}} constraints decrease model flexibility, increasing model error when the model is already highly inflexible. The UM-CEnKF estimating the restrictive Houlihan-selected parameters $P_{RH}$ increased forecast error for about $1/3^{rd}$ of patients. In contrast, the same constraints applied to the UM-CEnKF with the less restrictive standard Houlihan parameters $P_{H}$ reduced forecast errors. This effect is implicit model error, the UM-CEnKF estimating the $5$ $P_{RH}$-parameters may not be flexible enough to estimate patients and constraints make the problem worse.

\emph{\underline{Theme three:}} constraining \emph{observed} states does not positively impact forecast accuracy. Constraining glucose never reduces forecast errors but does not hurt forecasting errors as long as the constraints are not too severe; we deduce that data from natural and effective inference constraints.

\emph{\underline{Theme four:}} it was not difficult to identify effective and robust constraint boundaries and, in particular, to identify constraint structures which work well for the entire population. The constraint structure can have both a positive and negative impact on forecast accuracy, but assuming that the functional form of the model is flexible enough to approximate the data, constraints that reduced forecast error were not difficult to find.


\begin{table*}
\centering
\resizebox{\textwidth}{!}{
\begin{tabular}{|l| |l| |l| |l| |l| |l| |l| |l| |l| |l| |l| |l|}
  \hline
  \multicolumn{12}{|c|}{\textbf{UM$(P_{RH})$ (restrictive Houlihan): Ratio of CEnKF/EnKF forecast MSE after 24 data points ($<24$ hr)}} \\
  \hline \hline
patient& gm & gs & im & is & ius & gim & gis & gipm & gips & gpmius & gpsius \\ \hline
296 & 0.99&1.0&1.0&0.99&0.99&0.99&1.0&0.99&1.0&0.99&1.0  \\ \hline \hline
426 & 0.99&0.99&0.99&1.0&2.4&0.99&1.0&0.99&1.0&2.4&2.4  \\ \hline \hline
430 & 0.99&0.99&1.0&0.89&0.87&0.99&0.94&0.96&0.99&0.83&0.96  \\ \hline \hline
456 & 1.0&1.0&1.0&1.0&2.1&1.0&1.0&1.0&1.0&2.1&2.1 \\ \hline \hline
489 & 0.99&0.99&1.6&1.6&1.5&1.6&1.3&1.4&1.0&1.3&1.1  \\ \hline \hline
585 & 0.92&0.91&0.97&0.98&1.4&0.97&0.98&0.97&1.4&1.4&1.6  \\ \hline \hline
593 & 1.0&0.98&0.96&0.91&1.0&0.97&0.89&0.96&0.96&0.87&0.95  \\ \hline \hline
646 & 0.95&0.96&1.0&1.5&2.1&1.0&1.5&1.028&1.7&2.1&1.6  \\ \hline \hline
851 & 0.99&1.1&1.6&1.9&2.3&1.6&1.7&1.64&1.1&3.2&1.1   \\
\hline \hline
  \multicolumn{12}{|c|}{\textbf{Fraction of population with improved forecast with imposed constraints}} \\
  \hline \hline
harm: $\frac{MSE(CEnKF)}{MSE(EnKF)} \geq 1.1$ & 0&0.11&0.22&0.33&0.66&0.22&0.33&0.22&0.33&0.66&0.66 \\ \hline \hline
no change: $0.9 \leq \frac{MSE(CEnKF)}{MSE(EnKF)} <1.1$ & 1&0.88&0.77&0.55&0.22&0.77&0.55&0.77&0.66&0.11&0.33 \\ \hline \hline
mild improvement: $0.6 \leq \frac{MSE(CEnKF)}{MSE(EnKF)} <0.9$ & 0&0&0&0.11&0.11&0&0.11&0&0&0.22&0  \\ \hline \hline
substantial improvement: $\frac{MSE(CEnKF)}{MSE(EnKF)}<0.6$ & 0&0&0&0&0&0&0&0&0&0&0 \\ \hline \hline

   \hline
  \multicolumn{12}{|c|}{\textbf{UM$(P_{RH})$ (restrictive Houlihan) \emph{with bolus}: Ratio of CEnKF/EnKF forecast MSE after 24 data points ($<24$ hr)}} \\
  \hline \hline
patient& gm & gs & im & is & iuss & gim & gis & gipm & gips & gpmius & gpsius \\ \hline
489 & 0.99&0.99&1.6&1.6&1.5&1.7&1.3&1.4&1.0&1.3&1.1 \\ \hline \hline
585 & 0.91&0.91&0.96&0.98&1.4&0.96&0.97&0.96&1.4&1.4&1.6 \\ \hline \hline
851 & 0.98&1.0&1.6&1.8&2.3&1.6&1.8&1.7&1.1&3.7&1.4  \\ \cline{1-12}
\hline
  \multicolumn{12}{|c|}{\textbf{Fraction of population with improved forecast with imposed constraints}} \\
  \hline \hline
  harm: $\frac{MSE(CEnKF)}{MSE(EnKF)} \geq 1.1$ & 0&0&0.66&0.66&1.0&0.66&0.66&0.66&0.66&1.0&1.0 \\ \hline \hline
no change: $0.9 \leq \frac{MSE(CEnKF)}{MSE(EnKF)} <1.1$ & 1&1&0.33&0.33&0&0.33&0.33&0.33&0.33&0&0 \\ \hline \hline
mild improvement: $0.6 \leq \frac{MSE(CEnKF)}{MSE(EnKF)} <0.9$ & 0&0&0&0&0&0&0&0&0&0&0  \\ \hline \hline
substantial improvement: $\frac{MSE(CEnKF)}{MSE(EnKF)}<0.6$ & 0&0&0&0&0&0&0&0&0&0&0 \\ \hline \hline
   \hline \hline
\end{tabular}
}

\resizebox{\textwidth}{!}{
\begin{tabular}{|l| |l| |l| |l| |l| |l| |l| |l| |l| |l| |l| |l|}
  \hline
  \multicolumn{12}{|c|}{\textbf{UM$(P_H)$ (standard Houlihan): Ratio of CEnKF/EnKF forecast MSE after 24 data points ($<24$ hr)}} \\
  \hline \hline
patient& gm & gs & im & is & iuss & gim & gis & gipm & gips & gpmius & gpsius \\ \hline
296 & 1.0&1.0&1.0&1.0&1.4&1.0&1.0&1.0&1.0&1.4&1.4  \\ \hline \hline
426 & 1.0&1.0&1.0&1.0&1.7&1.0&1.0&1.0&1.0&1.7&1.7  \\ \hline \hline
430 & 1.0&1.0&1.3&0.73&0.63&1.3&0.73&1.3&5.6&0.61&5.8  \\ \hline \hline
456 & 1.0&1.0&1.0&1.0&1.8&1.0&1.0&1.0&1.0&1.8&1.8  \\ \hline \hline
489 & 1.0&1.0&0.74&0.4&0.48&0.74&0.41&0.78&1.3&0.44&1.1  \\ \hline \hline
585 & 1.0&1.0&1.0&1.0&2.7&1.0&1.0&1.0&1.1&2.8&2.8 \\ \hline \hline
593 & 1.1&1.1&0.21&0.21&0.71&0.21&0.21&0.21&3.8&0.71&4.0 \\ \hline \hline
646 &1.0&1.0&1.0&0.82&0.68&1.0&0.82&1.0&0.93&0.68&1.2  \\ \hline \hline
851 & 1.0&1.0&0.61&0.43&0.55&0.61&0.43&0.63&1.6&0.46&1.9  \\ \cline{1-12}
\hline
  \multicolumn{12}{|c|}{\textbf{Fraction of population with improved forecast with imposed constraints}} \\
  \hline \hline
  harm: $\frac{MSE(CEnKF)}{MSE(EnKF)} \geq 1.1$ & 0&0.11&0.11&0&0.44&0.11&0&0.11&0.55&0.44&1.0 \\ \hline \hline
no change: $0.9 \leq \frac{MSE(CEnKF)}{MSE(EnKF)} <1.1$ & 1&0.88&0.55&0.44&0&0.55&0.44&0.55&0.44&0&0 \\ \hline \hline
mild improvement: $0.6 \leq \frac{MSE(CEnKF)}{MSE(EnKF)} <0.9$  & 0&0&0.22&0.22&0.33&0.22&0.22&0.22&0&0.33&0  \\ \hline \hline
substantial improvement: $\frac{MSE(CEnKF)}{MSE(EnKF)}<0.6$ & 0&0&0.11&0.33&0.22&0.11&0.33&0.11&0&0.22&0 \\ \hline \hline

   \hline
  \multicolumn{12}{|c|}{\textbf{UM$(P_H)$ (standard Houlihan) \emph{with bolus}: Ratio of CEnKF/EnKF forecast MSE after 24 data points ($<24$ hr)}} \\
  \hline \hline
patient& gm & gs & im & is & iuss & gim & gis & gipm & gips & gpmius & gpsius \\ \hline
489 & 1.0&1.0&0.75&0.41&0.48&0.75&0.42&0.76&1.3&0.43&1.0 \\ \hline \hline
585 & 1.0&1.0&1.0&1.0&2.8&1.0&1.0&1.0&1.1&2.8&2.8 \\ \hline \hline
851 & 1.1&1.2&0.60&0.45&0.58&0.60&0.45&0.61&1.5&0.49&1.7  \\ \cline{1-12}
\hline
  \multicolumn{12}{|c|}{\textbf{Fraction of population with improved forecast with imposed constraints}} \\
  \hline \hline
  harm: $\frac{MSE(CEnKF)}{MSE(EnKF)} \geq 1.1$ & 0&0.33&0&0&0.33&0&0&0&1.0&0.33&0.66 \\ \hline \hline
no change: $0.9 \leq \frac{MSE(CEnKF)}{MSE(EnKF)} <1.1$ & 1&0.66&0.33&0.33&0&0.33&0.33&0.33&0&0&0.3333 \\ \hline \hline
mild improvement: $0.6 \leq \frac{MSE(CEnKF)}{MSE(EnKF)} <0.9$ & 0&0&0.66&0&0&0.66&0&0.66&0&0&0 \\ \hline \hline
substantial improvement: $\frac{MSE(CEnKF)}{MSE(EnKF)}<0.6$ & 0&0&0&0.66&0.66&0&0.66&0&0&0.66&0 \\ \hline

\end{tabular}
}
 \caption{Results from the CEnKF constraint experiments: we report the ratio of $\omega=\frac{MSE(CEnKF)}{MSE(EnKF)}$ where $\omega \geq 1.1$ indicates the forecast was worsened by constraints, $1.1 > \omega  \geq 0.9$ indicates the forecast was not impacted by constraints, $0.9 > \omega \geq 0.6$ indicates the forecast was improved by constraints, and $0.6 > \omega $ indicates the forecast was substantially improved by constraints. \emph{Chief result} constraining the unmeasured states are the most effective constraints; here severely constraining insulin never hurts and otherwise improves forecast accuracy.}
   \label{table:constraint_mse_ratio}
\end{table*}

\section{Discussion}

\paragraph{\textbf{Overarching results:}} We developed and implemented a computational pipeline for using the CEnKF with clinical, here ICU blood glucose, data to forecast physiological features. We show that the constraints applied to the EnKF make it more robust and lead to more plausible physiological forecasts for more patients with less data compared with the unconstrained EnKF. We show that the choice of parameters to estimate can have a substantial impact on the stability and success of physiological forecasting with DA: if we use too few parameters, the model is too rigid to estimate the data well, and if we use too many parameters, identifiability failure makes forecasting non-robust. And most surprisingly, while the CEnKF arrived at parameter values that minimized the least squared error with less data, it also led to glucose dynamics with attenuated oscillations; thus, while these parameters optimize a least squares objective, they may also be physiologically incorrect, as regular glucose oscillations are known to occur in ICU patients constant nutritional feeding.

\paragraph{\textbf{Three cardinal conclusions, given our results}} \emph{Our first major conclusion}: DA can be applied with ICU (EHR) data to forecast physiological processes, which is important given the complexity of operationally collected clinical data. However, the forecasts are not yet robust for all patients. \emph{Our second major conclusion}: getting DA to work given data-sparsity and time-sensitive limitations requires limiting the model space. However, it is possible to limit the model too much and render it too simple to allow for accurate representation of the data. To produce accurate forecasts, there is an optimal balance between model complexity and model simplicity that must be found. This balance can be difficult to find and often requires multiple approaches. \emph{Our third major conclusion}: both constraining the EnKF and selecting parameters with the Houlihan method significantly reduce data required to achieve usefully low forecast error. We find that constraints improve forecast accuracy significantly when mild to severe constraints are applied to states and parameters that are not measured. We find there is a balance between over- and under-constraining states and parameters---this is not surprising and is similar in flavor to regularization. Over-constraining effectively turns off inference of a given parameter and does not allow the model the flexibility to accurately predict patients' glycemic response, given real data. Under-constraining allows the model too much flexibility and accurate, convergent solutions cannot be found with sparse data.

\paragraph{\textbf{Using DA to forecast physiological features with clinical sparse data}} There are two approaches for using sparse data in an ICU to forecast personal physiological state: (i) constrain the model through constrained inference, or (ii) constrain the model space by simplifying the model. Here we explored a constrained inference approach. We demonstrated that while simple constraints make forecasting more robust with less data, constrained inference does not completely solve the data sparsity problem. In particular, there can be many plausible least squares minima that lead to model dynamics that do not represent the correct underlying physiological dynamics. Therefore, it will be important to explore the use of different metrics other than least squares, more sophisticated constraints that retain important physiological trajectory dynamics, and new DA methods that manage errors in measurement timing. The other approach, development of simplified, identifiable models is also an important research direction. A new class of physiologic models are needed that balance biological realism with practical identifiability subject to real world data. In the context at hand, the minimal stochastic glucose (MSG) model \cite{melike2019simple} is a one-dimensional model of glucose dynamics that replaces the never-measured, hard-to-identify insulin state with a parametric representation. This model is far less sophisticated than the Ultradian model used here, yet captures some key dynamical features of glucose homeostasis and response to external insulin stimuli. The MSG model loses some physiological fidelity and accuracy, e.g., stochastic models do not have explicit blood glucose trajectories. But such models exchange fidelity and trajectory accuracy for gains in forecast robustness and potentially gains in the interpretability of the lumped parameters. Because measuring humans always comes with a cost and the humans we need to predict tend to become nonstationary effectively making the data sparse, both approaches need to be explored.

\paragraph{\textbf{Forecast accuracy and evaluation as applied to clinical settings}} We observed several complex issues related to forecast accuracy and its evaluation. \emph{First}, the DA often under-estimated the measured extrema, usually on the upper end of the blood glucose distribution. Simultaneously achieving accuracy of the extrema of physiological dynamics and estimating small-scale oscillations with sparse data is an interesting challenge. \emph{Second}, MSE, linear correlation and distribution-based metrics by themselves do not help evaluate impact of either forecast accuracy for decision-making or quantities, e.g., glycemic distribution boundaries, that are important for practical use of the DA-produced information. While there exist error metrics that address this issue, e.g., the Parkes error grid \cite{parkes_grid,george_eval}, measures that include both clinical consequence and numerical error are rare. In this way, the evaluation itself is difficult without advances in skill scores \cite{forecast_verification} more tailored to the needs of biomedicine or clinical decision-making.

\paragraph{\textbf{Lessons regarding the use of constrained DA to forecast physiological features}} We have four conclusions concerning the application of constraints to the EnKF to estimate physiology using ICU data. \emph{First}, constraining \emph{unmeasured} states and parameters mitigates identifiability failure due to lack of measurements and substantially reduces estimation error. \emph{Second}, constraining measured states generally does not reduce estimation error. It is better to allow the data to constrain the inference, although there is a question about whether this is true in the case of severe outliers. \emph{Third}, setting constraint boundaries does not require exacted, perfect knowledge to gain the advantages from the constraints; it is enough to use outside knowledge of how the natural system should behave, e.g., insulin between $75-275$. It is likely that using machine learning to fine-tune constraints would result in further estimation error reduction. And \emph{fourth}, applying constraints to models that are too rigid to estimate the data further worsens estimation errors.


\paragraph{\textbf{The impact of sparse data on Kalman-type inference}}  We often observe the UM-EnKF and UM-CEnKF underestimating the variance of the data distribution. Figure \ref{fig:593umh2p50_c_is} reveals a potential reason: the UM-CEnKF ensemble trajectories do not well match the oscillatory frequencies present in the data and in many cases are effectively approximating mean homeostasis given nutrition. We hypothesize that the source of this problem is \emph{a dynamics mismatch problem} due to data sparsity in combination with the $l_2$-minimization within the Kalman filter inference methodology. \emph{Notably, nonlinear Kalman filters paired with nonlinear models can outperform linear filter-model pairs when the generating system is nonlinear, but this is only possible if they have the data to resolve the nonlinear dynamics.} In the context of this paper where we use ICU data that are sparse relative to the oscillatory dynamics, the sparsity raises the question: \emph{how sparsely measured can these data be before the solution quality of the linear and nonlinear filter-model pairs coincide?} For example, there could be a nonlinear filter-model pair that cannot turn off oscillations and instead aliases---so it will be ``bad" in an MSE sense, but also ``different" from the linear solution.

The intuition for this problem follows along the following reasoning. The optimal solution for the Kalman (linear) filter is the conditional mean of the data \cite{filtering_jaz}. Moreover, linear filters with linear models estimate parameters that yield a single optimal solution. In contrast, nonlinear filter-model pairs do not have identifiable optimal solutions and can have many local minima that yield very different model dynamics yet these solutions can be difficult to differentiate by least squared error alone. When data are densely sampled in time and space, the nonlinear filter can exploit nonlinear/higher-order information within more complex nonlinear models to make predictions and validate scientific hypotheses. As data become more sparse, it becomes more difficult to resolve and exploit nonlinear/higher-order information and a problem emerges: the nonlinear filter must select the set of parameters for the model it is paired with that best synchronize the model with the generating data \emph{by minimizing the least squared distance between the model output and the data from many possible local minima.}  Given the many possible parameter minima, the solution we want is one that minimizes \emph{two properties, the least square minimizer and the dynamics mismatch for which there is not an obvious minimizer.} It is not difficult to construct sparsely measured examples where many of the local minima of a nonlinear model correspond to a relatively low MSE \emph{while the model parameters lead to dynamics that do not represent the underlying dynamics of the generating system.} In many of these examples the conditional mean solution for the sparsely sampled data may correspond to parameters of the forward model whose equilibrium is a fixed point leading to a dynamics mismatch with the sparsely sampled oscillatory generating system. The transition from dense to sparse data implies a data sparsity driven turning point where simpler model-filter pair---e.g., linear models and linear stochastic models \cite{melike2019simple} estimated with a Kalman filter---will become more robust, accurate and useful for making predictions and validating scientific hypotheses than the nonlinear models paired with nonlinear filters that more closely resemble the generating system. We think there is likely an inflection point in the sampling rate where the measure of the set of nonlinear model parameters found to be good solutions by the nonlinear filters, specified by MSE, will become proportionally smaller than the set of linear model parameters found to be good solutions by the linear filters, which leads to a conjecture:



\emph{\textbf{Data assimilation sampling conjecture:} There exists a sampling rate, e.g., the Nyquist rate, of sampling data above which there will be a nonlinear model-filter pair that can outperform any linear model-filter pair (whose solution is the conditional mean of the data, is a fixed point). As the sampling rate decreases below this rate it becomes increasingly probable, with probability approaching one, that there will be a linear model-filter pair that will be the optimal solution for all Kalman-type filters.}


Proving such a conjecture is difficult. A path forward is to understand the dynamics mismatch problem, meaning, devising an evaluation or model selection methodology that is sensitive to (i) the least squares distance between the model output and data, and (ii) additional information regarding the underlying generating and model dynamics.

\paragraph{\textbf{Limitations:}} Our study has several limitations. The data set we use for this paper is very accurate, it was manually curated, but it is small and not expansive. Specifically, we did not include medications that can impact glyemic dynamics and confound inference such as glucocorticoids. We included patients that only had two of the four broad types of insulin---we included rapid- and short-acting among the possibilities of rapid-acting, short-acting, intermediate-acting, long-acting. We did not vary DA methods beyond ensemble Kalman filter techniques. It is possible that including, e.g., particle filters or optimization methods, would impact our results. And finally, we did not vary the physiological models as we did in, e.g., \cite{albers_plos_comp_bio_DA_I}, to test whether models with different parameter sets and different underlying dynamics would behave differently.

\paragraph{\textbf{A path forward}} \emph{First}, estimating parameter initial conditions to retain qualitatively important features, e.g., oscillation amplitude, mean, etc., restricted data collected in the first $12-24$ hours will likely lead to more accurate parameter estimation and tracking with less data. \emph{Second}, parameter inference as it is currently implemented is not guaranteed to make the model more representative of the system dynamics. \emph{Third}, there is potentially useful information in trustable parameter trends---they indicate the patient health state trajectory. Computing verifiably trustable parameter trajectories would be a significant advance. Specifically, progress toward updating parameters only when the update improves over random fluctuation and more uncertainty quantification of parameters would represent significant advancements. \emph{Fourth}, model development that accommodates realistic data and includes clinically meaningful parameters without overwhelming the system with identifiability issues would move the usefulness of DA forward a great deal. This effort may be aided by hierarchical approaches that aggregate patient-level data sets to improve individual predictions. And \emph{fifth}, devising skill scores and other evaluation metrics that reflect the useful aspects of forecasts and parameter estimates in the context of biomedicine or clinical decision-making would be long-lasting contributions.

\paragraph{\textbf{Acknowledgements:}} We would like to thank Andrew Stuart for many helpful discussions and reading the manuscript, and to Jacob Stroh, Tellen Bennett, Cecilia Low Wang, William Ott and Bhargav Karamched for many helpful discussions related to this work.


\clearpage
\appendix

\section{Ultradian model}
\label{app:ultradian}

The model is comprised of a set of six ordinary differential equations; the model is non-autonomous because it has an external, time-dependent driver, consumed nutrition. The six dimensional state space made up of three physiologic variables and a three stage filter.  The physiologic state variables are the glucose concentration $G$, the plasma insulin concentration $I_{p}$, and the interstitial insulin concentration $I_{i}$. The three stage filter $(h_1,h_2,h_3)$ which reflects the response of the plasma insulin to glucose levels \cite{sturis_91}. The model was designed to capture ultradian oscillations missing in previous models. The ordinary differential equations that define the model are \cite{keenerII}:
\begin{eqnarray}
\label{eq:model1}
\frac{dI_p}{dt} & = &  f_1(G)-E\bigl(\frac{I_{p}}{V_{p}}-\frac{I_i}{V_{i}}\bigr)-\frac{I_{p}}{t_{p}}\\
\frac{dI_i}{dt} & = & E\bigl(\frac{I_{p}}{V_{p}}-\frac{I_i}{V_{i}}\bigr)-\frac{I_{i}}{t_{i}}\\
\frac{dG}{dt} & = & f_4(h_3)+I_{G}(t)-f_2(G)-f_3(I_i)G\\
\frac{dh_1}{dt} & = & \frac{1}{t_d}\bigl(I_p-h_1\bigr) \\
\frac{dh_2}{dt} & = & \frac{1}{t_d}\bigl(h_1-h_2\bigr) \\
\frac{dh_3}{dt} & = & \frac{1}{t_d}\bigl(h_2-h_3\bigr)
\end{eqnarray}


The state variables include physiologic processes that have been parameterized, including: $f_1(G)$ represents the rate of insulin production; $f_2(G)$ represents insulin-independent glucose utilization; $f_3(I_i)G$ represents insulin-dependent glucose utilization; $f_4(h_3)$ represents delayed insulin-dependent glucose utilization.  These functions are defined by:
\begin{eqnarray}
f_1(G) & = & \frac{R_m}{1+ \exp(\frac{-G}{V_g c_1} + a_1)} \\
f_2(G) & = & U_b(1-\exp(\frac{-G}{C_2V_g})) \\
f_3(I_i) & = & \frac{1}{C_3 V_g}( U_0 + \frac{U_m - U_0}{1 + (\kappa I_i)^{-\beta}}) \\
f_4(h_3) & = & \frac{R_g}{1 + \exp(\alpha (\frac{h_3}{C_5 V_p}  -1))} \\
\kappa & = & \frac{1}{C_4} (\frac{1}{V_i} - \frac{1}{E t_i})
\end{eqnarray}

The nutritional driver of the model $I_G(t)$ is defined over $N$ discrete nutrition events \cite{dyn_pheno}, where $k$ is the decay constant and event $j$ occurs at time $t_j$ with carbohydrate quantity $m_j$
\begin{equation}
\label{eq:ig}
I_G(t) = \sum^N_{j=1} \frac{m_j k}{60} \exp(k(t_j-t));  N=\# \{ t_j < t \}
\end{equation}

\begin{table*}[!ht]
\centering
\caption{Full list of parameters for the ultradian glucose-insulin model \cite{keenerII}. Note that IIGU and IDGU denote insulin-independent glucose utilization and insulin-dependent glucose utilization, respectively.}
\begin{tabular}{|p{1.2cm}|p{2.7cm}|p{8cm}|}
\hline
\multicolumn{3}{|p{8cm}|}{\textbf{Ultradian model parameters}} \\ \hline
\Cline{2pt}{1-3} \hline
\Cline{2pt}{1-3}
Name & Nominal Value  & Meaning \\ \hline \hline
$V_p$  & $3$ l  & plasma volume  \\ \hline \hline
$V_i$  & $11$ l  & interstitial  volume \\ \hline \hline
$V_g$ & $10$ l  & glucose space \\ \hline \hline
$E$  & $0.2$ l min$^{-1}$ &   exchange rate for insulin between remote
and plasma compartments \\ \hline \hline
$t_p$  & $6$ min  & time constant for plasma insulin degradation (via
kidney and liver filtering) \\ \hline \hline
$t_i$  & $100$ min & time constant for remote insulin degradation (via
muscle and adipose tissue) \\ \hline \hline
$t_d$  & $12$ min  & delay between plasma insulin and glucose
production \\ \hline \hline
$k$  & $0.5$ min$^{-1}$  & rate of decayed appearance of ingested glucose \\ \hline \hline
$R_m$  & $209$ mU min$^{-1}$  & linear constant affecting insulin secretion  \\ \hline \hline
$a_1$  & $6.6$ & exponential constant affecting insulin secretion \\ \hline \hline
$C_1$  & $300$ mg l$^{-1}$ & exponential constant affecting insulin secretion \\ \hline \hline
$C_2$  & $144$ mg l$^{-1}$  & exponential constant affecting IIGU \\ \hline \hline
$C_3$  & $100$ mg l$^{-1}$  & linear constant affecting IDGU \\ \hline \hline
$C_4$  & $80$ mU l$^{-1}$ & factor affecting IDG \\ \hline \hline
$C_5$  & $26$ mU l$^{-1}$  & exponential constant affecting IDGU \\ \hline \hline
$U_b$  & $72$ mg min$^{-1}$  & linear constant affecting IIGU \\ \hline \hline
$U_0$  & $4$ mg min$^{-1}$ & linear constant affecting IDGU \\ \hline \hline
$U_m$  & $94$ mg min$^{-1}$  &  linear constant affecting IDGU \\ \hline \hline
$R_g$  & $180$ mg min$^{-1}$  & linear constant affecting IDGU \\ \hline \hline
$\alpha$  & $7.5$ & exponential constant affecting IDGU \\ \hline \hline
$\beta$  & $1.772$ & exponent affecting IDGU \\ \hline \hline
\end{tabular}
\label{table:model_parameters}
\end{table*}

\bibliographystyle{elsarticle-num}
\bibliography{master,dstexts,srb,structuralstability,partialhyperbolicity,neuralnetworks,bifurcationtheory}

\end{document}